\documentclass[]{JHEP3}
\JHEPspecialurl{http://jhep.sissa.it/JOURNAL/JHEP3.tar.gz}

\usepackage{graphicx}
\usepackage{amsmath}
\usepackage{bbm}
\usepackage{url}

\newcommand{\be}{\begin{equation}}
\newcommand{\ee}{\end{equation}}
\newcommand{\bea}{\begin{eqnarray}}
\newcommand{\eea}{\end{eqnarray}}
\newcommand{\Eq}[1]{Eq.~(\ref{#1})}

\newcommand{\Sec}[1]{Sec.~\ref{#1}}
\newcommand{\App}[1]{App.~\ref{#1}}
\newcommand{\Fig}[1]{Fig.~\ref{#1}}
\newcommand{\Ref}[1]{Ref.~\cite{#1}}
\newcommand{\Tab}[1]{Table~\ref{#1}}

\newcommand{\vev}[1]{\langle #1 \rangle}
\newcommand{\CO}{\mathcal{O}}
\newcommand{\TeV}{\text{TeV}}

\newcommand{\CW}{\rm CW}
\newcommand{\tr}{\mathop{\rm tr}}
\newcommand{\diag}{\mathop{\rm diag}}
\renewcommand{\Re}{\mathop{\rm Re}}
\renewcommand{\Im}{\mathop{\rm Im}}

\title{A Fat Higgs with a Magnetic Personality}

\author{Nathaniel Craig\\
	School of Natural Sciences, Institute for Advanced Study,\\
	Princeton, NJ 08540 \\
	Department of Physics and Astronomy, Rutgers University, \\
	Piscataway, NJ 08854 \\
	E-mail: \email{ncraig@ias.edu}}
	
\author{Daniel Stolarski\\
	Center for Fundamental Physics, Department of Physics, University of Maryland,\\ 
	College Park, MD 20742\\
	Department of Physics and Astronomy, Johns Hopkins University,\\
	Baltimore, MD 21218\\
	E-mail: \email{danchus@umd.edu}}
	
\author{Jesse Thaler\\
	Center for Theoretical Physics, Massachusetts Institute of Technology, \\
	Cambridge, MA 02139 \\
E-mail: \email{jthaler@jthaler.net}}

\preprint{RUNHETC-2011-13 \\ UMD-PP-11-007 \\ MIT-CTP 4271}

\abstract{We introduce a novel composite Higgs theory based on confining supersymmetric QCD.  Supersymmetric duality plays a key role in this construction, with a ``fat'' Higgs boson emerging as a dual magnetic degree of freedom charged under the dual magnetic gauge group.  Due to spontaneous color-flavor locking in the infrared, the electroweak gauge symmetry is aligned with the dual magnetic gauge group, allowing large Yukawa couplings between elementary matter fields and the composite Higgs.  At the same time, this theory exhibits metastable supersymmetry breaking, leading to low-scale gauge mediation via composite messengers.  The Higgs boson is heavier than in minimal supersymmetric theories, due to non-decoupling $D$-terms and a large $F$-term quartic coupling.  This theory predicts quasi-stable TeV-scale pseudo-modulini, some of which are charged under standard model color, possibly giving rise to long-lived $R$-hadrons at the LHC.}

\keywords{Beyond Standard Model, Technicolor and Composite Models, Supersymmetric Standard Model}

\begin{document} 

\section{Introduction}

The origin of electroweak symmetry breaking (EWSB) is a central question in and beyond the standard model (SM).  At one extreme, EWSB could be triggered by the vacuum expectation value (vev) of an elementary Higgs scalar whose potential may or may not be stabilized by supersymmetry (SUSY).  At the other extreme, EWSB could be triggered by the vev of a composite operator as in technicolor theories.  Here, we wish to revisit the intermediate possibility that a Higgs scalar might emerge as a composite state from strong dynamics.   We will use supersymmetric duality as a tool to build a realistic model where the Higgs is a magnetic degree of freedom from confining electric dynamics.   

There are a variety of reasons to suspect that physics at the electroweak scale might be described by a supersymmetric but composite theory.  Weak scale SUSY is an attractive solution to the hierarchy problem since it stabilizes the electroweak scale without introducing large corrections to precision electroweak observables.  On the other hand, the non-observation of the Higgs boson at LEP points towards additional dynamics beyond the minimal supersymmetric standard model (MSSM) to raise the Higgs boson mass.   Composite or ``fat'' Higgs theories \cite{Harnik:2003rs} enable the Higgs to have stronger self-couplings than typically allowed in perturbative SUSY scenarios, raising the physical Higgs mass while still preserving many of the desired features of weak scale SUSY.\footnote{While generic SUSY composite models do not exhibit gauge coupling unification, we will take the attitude that the virtues of compositeness outweigh the loss of manifest unification.}

In this paper, we introduce a new type of SUSY composite Higgs theory based on confining supersymmetric QCD (SQCD) with $N_c$ colors and $N_f = N_c + 2$ fundamental flavors in the ultraviolet (UV), and a compositeness scale $\Lambda \simeq 1000~\TeV$. The simplest case with calculable infrared (IR) dynamics corresponds to $N_f = 7$, $N_c = 5.$ This model has a number of unique features.
\begin{itemize}
\item The Higgs superfields are identified with dual quark fields, meaning that they are composite degrees of freedom with no (simple) UV interpolating operators.  While previous fat Higgs models have utilized composite meson \cite{Harnik:2003rs, Chang:2004db} or baryon \cite{Delgado:2005fq} fields, to our knowledge this is the first time a SM mode has been identified with a dual quark field emerging at relatively low energies.\footnote{Models in which all SM fields are dual degrees of freedom from GUT-scale duality were first constructed in, e.g. \Ref{Berkooz:1997bb}. In contrast, here the scale of duality is low and the magnetic dynamics play a crucial role.} 
\item In order for the Higgs bosons to have the correct electroweak quantum numbers, $SU(2)_L$ must be aligned with the dual magnetic group.  That is, even the transverse SM $W$ and $Z$ bosons are partially composite states.  Again, to our knowledge this is a novel use for a magnetic gauge group.
\item The Higgs boson can be heavier than in the MSSM because the magnetic gauge group leads to a non-decoupling $D$-term.  In addition, there is a singlet meson field that produces an additional NMSSM-like Higgs quartic couplings. This coupling may be naturally quite large, as duality provides a well-behaved UV completion at high energies. 
\item To ensure the existence of a dual magnetic group, $N_f$ must fall in the range $\frac{3}{2}N_c > N_f >  N_c$.  This turns out to be the same range for which SQCD exhibits metastable SUSY breaking.  Thus, SUSY breaking is automatically tied to Higgs compositeness in this model, and it is natural to have a modified version of direct gauge/gaugino mediation.
\item Despite the fact that the Higgs bosons are dual squark fields charged under a dual magnetic group, one can still achieve a large Yukawa coupling to an elementary top quark.  This is possible because of the color-flavor locking phenomenon in SQCD combined with the technique of ``bosonic technicolor'' \cite{Samuel:1990dq,Dine:1990jd}.
\item While some features of composite Higgs theories can only be estimated through naive dimensional analysis, the existence of a weakly-coupled magnetic dual of a confining electric theory gives us important calculational handles to assess the viability of our scenario.  
\end{itemize}
Together, these features lead to an intriguing paradigm where a single dynamical sector leads to both EWSB and SUSY breaking in a calculable regime.\footnote{For related constructions connecting supersymmetric strong dynamics, EWSB, and SUSY breaking (albeit without magnetic gauge fields), see also \Ref{Luty:2000fj, Murayama:2003ag, SchaferNameki:2010iz, SchaferNameki:2010mg, Fukushima:2010pm}.}  The scales that feature in this scenario are summarized in \Fig{fig:Scales}.

\FIGURE[t]{
\includegraphics[scale=0.7]{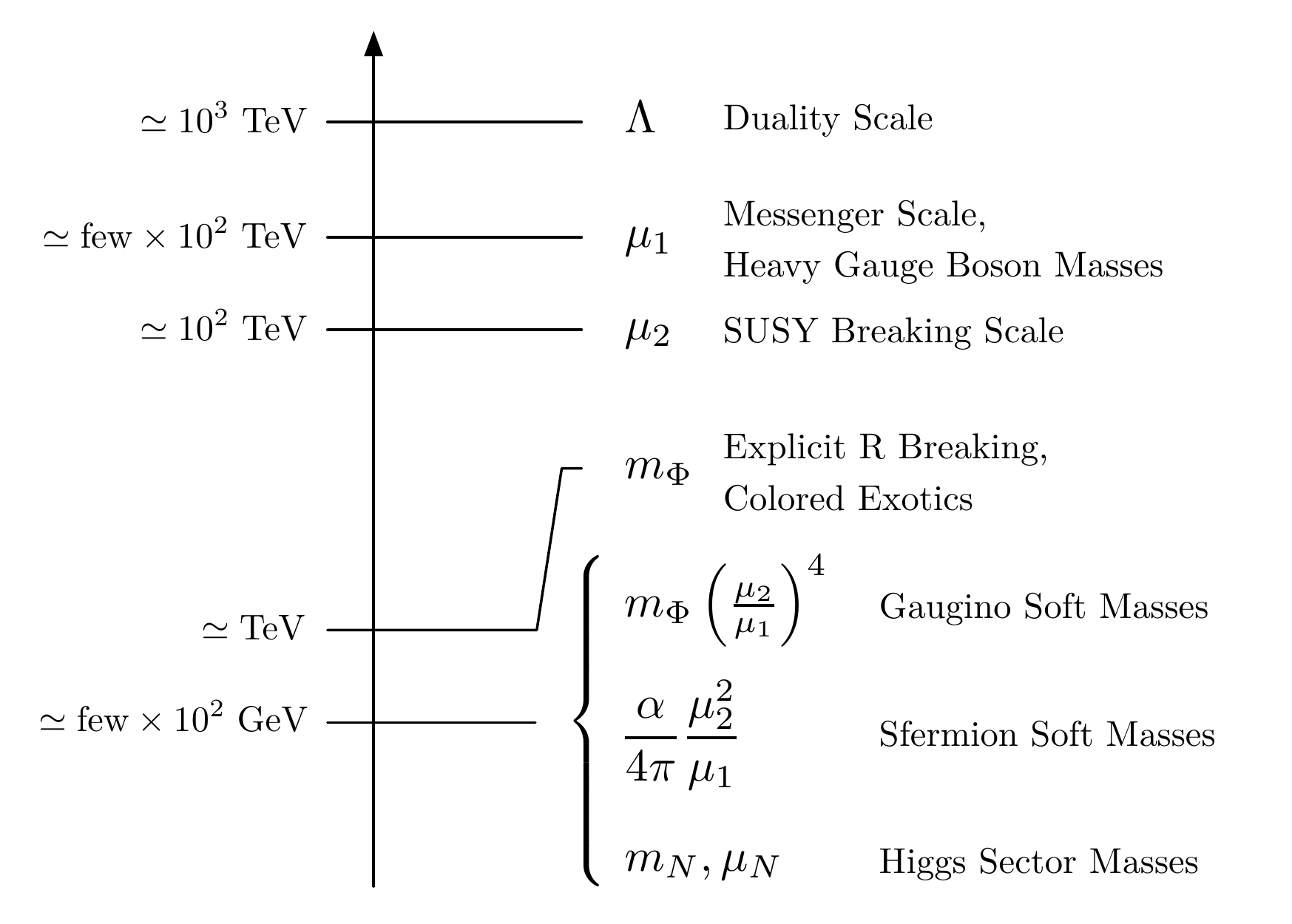}
\caption{Sketch of the approximate scales that arise in our construction. The electric theory grows strong at a scale $\Lambda \gtrsim 10^3$ TeV, below which SUSY and electroweak symmetry are broken around $10^2$ TeV and $10^2$ GeV, respectively. The details of these mass scales  are presented in \Sec{sec:theory}.   }
\label{fig:Scales}
}

The remainder of the paper is organized as follows.  We describe the UV field content and the resulting dual IR theory in \Sec{sec:theory}.  This model naturally incorporates both SUSY breaking and EWSB, described separately in \Sec{sec:SUSYbreaking} and \Sec{sec:EWSB}.  We highlight some of the main phenomenological features in \Sec{sec:pheno}, and conclude in \Sec{sec:conclude}.  Various minutiae are left to the appendices.   In \App{app:75} we render a more detailed picture of the minimal model with $N_f = 7, N_c = 5$. In \App{app:64} we consider the unusual case of $N_f = 6, N_c = 4,$ for which there is an additional marginal contribution to the superpotential from nonperturbative dynamics. Finally, in \App{app:soft} we make a detailed accounting of the states for general $N_f$.

\section{A Magnetic Composite Higgs}
\label{sec:theory}

The Higgs sector of any supersymmetric standard model (SSM) is special, since $H_u$ and $H_d$ form a vector-like pair.  It is therefore natural to generate composite Higgs states from a vector-like confining theory, while leaving the quarks and leptons of the SSM as elementary fields.  Here we will describe how Higgs multiplets can emerge as dual quark fields from SQCD, while still having large Yukawa couplings to SSM matter fields.

\subsection{The Electric Theory}

Our starting point is $SU(N_c)$ SQCD with $N_c$ colors coupled to $N_f = N_c + 2$ flavors of fundamental and antifundamental quarks $Q, \overline Q$. In the absence of any superpotential deformations, there is an $SU(N_f)_L \times SU(N_f)_R \times U(1)_V$ global symmetry.  This theory grows strong at a scale $\Lambda$, below which it may be described in terms of a dual $SU(N \equiv N_f - N_c = 2)$ magnetic gauge theory with $N_f$ flavors of dual quarks $q, \overline q$ and a dual meson $M$ transforming as a bifundamental of $SU(N_f)_L \times SU(N_f)_R$ \cite{Seiberg:1994pq}. This dual theory is weakly coupled provided $N_f < \frac{3}{2} N_c$, meaning that the smallest theory with calculable IR dynamics corresponds to $N_f = 7, N_c = 5$.\footnote{The theory with $N_f = \frac{3}{2} N_c$---i.e., $N_f = 6, N_c = 4$---is likewise free in the IR, but the dynamics are altered by the presence of a marginal nonperturbative superpotential. We will consider this special case in \App{app:64}, but otherwise will focus on the somewhat larger theories with $N_f < \frac{3}{2} N_c$ for which the nonperturbative superpotential is strictly irrelevant.} In what follows we will retain general values of $N_f$, and a detailed treatment of $N_f = 7, N_c = 5$ is reserved for \App{app:75}.

We will describe the magnetic dual in further detail in \Sec{sec:magnetictheory}, but first we need to add a number of deformations to this SQCD theory.  Apart from the anomalous axial $U(1)$, it is easiest to visualize the global symmetries as being $U(N_f)_L \times U(N_f)_R$.  The deformations will explicitly break this $U(N_f)_L \times U(N_f)_R$ flavor symmetry to a diagonal subgroup of $U(N_f)_D$.  The most important deformation treats one of the flavors as special since it will correspond to the Higgs states, and we call this flavor $P,\overline{P}$.  Thus, the symmetries of the theory are best understood in terms of $\left[U(1) \times U(N_f-1) \right]^2$ and the diagonal subgroup $U(1)_D \times U(N_f-1)_D$.  In \Fig{fig:UVMoose}, we summarize the UV field content in moose notation.   

\FIGURE[t]{
\includegraphics[scale=0.7]{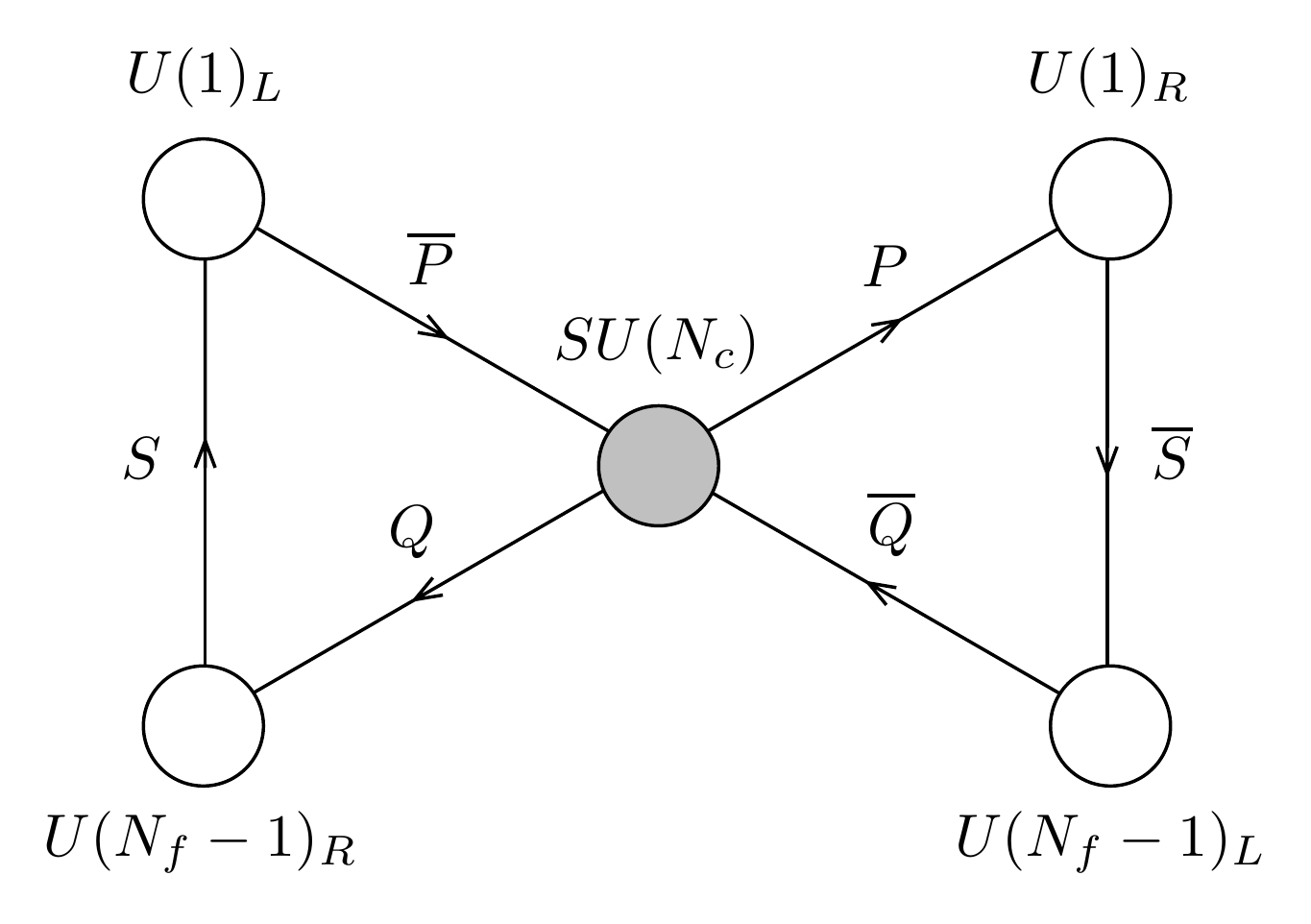}
\caption{The UV field content in moose notation.  The theory is an $SU(N_c)$ gauge theory with $N_f = N_c + 2$ flavors.  One of the flavors ($P, \overline{P}$) is singled out because it will be related to the Higgs multiplets in the IR.  Here, we are making explicit the $\left[U(1) \times U(N_f -1) \right]^2$ flavor quantum numbers, though quark mass terms will break the global symmetries to the diagonal subgroup $U(1)_D \times U(N_f -1)_D$.  Eventually, we will weakly gauge an unbroken diagonal subgroup of the flavor symmetries.  Note that the labels $L$ and $R$ are flipped for $U(N_f-1)$ in order to more easily draw in the spectator $S$, $\overline{S}$ fields.}
\label{fig:UVMoose}
}

 \TABLE[t]{
 \parbox{5in}{
 \begin{center}
\begin{tabular}{|c|c|cc|c|}
\hline
 & $SU(N_c)$ & $U(1)_D$ & $U(N_f-1)_D$ & $U(1)_V$  \\ \hline
$P$ & ${\bf \square}$ & $-1$& ${\bf 1}$ & $- 1/3$ \\
$\overline P$ & ${\bf \overline \square}$& $+1$ & ${\bf 1}$ & $+1/3$ \\
$Q$ & ${\bf \square}$ & 0 & ${\bf \overline \square}$ & $-1/3$ \\
$\overline Q$ & $ {\bf \overline \square}$ & $0$ & $\bf \square$ & $+1/3$ \\ \hline
$S$ & ${\bf 1}$ & $-1$& $\bf \square$ & $0$ \\
$\overline S$ & ${\bf 1}$ & $+1$ &  ${\bf \overline \square}$ & $0$ \\
\hline
\end{tabular}
\end{center}
}
\caption{The UV field content.  Unlike in \Fig{fig:UVMoose}, here we only give the quantum numbers under the diagonal $U(1)_D \times U(N_f-1)_D$ flavor symmetry.  For later convenience, we have identified the electric antiquark number symmetry $U(1)_V \subset U(N_f)_D$. \label{tab:UVfieldcontent}}
}
 
In \Tab{tab:UVfieldcontent}, we have organized the same fields according to their transformation properties under the diagonal $U(1)_D \times U(N_f-1)_D$ subgroup of the global symmetries.  We have also identified $U(1)_V \subset U(N_f)_D$ as electric antiquark number\footnote{In the IR, this symmetry will correspond to magnetic quark number.} with the generator
\be
V = \frac{1}{3} \, \diag(\underbrace{1,\ldots,1}_{N_f}).
\label{eq:Vgen}
\ee
We are using a notation for the diagonal $U(N_f)_D$ where the first entry corresponds to $U(1)_D$ (the $P,\overline{P}$ quarks) and the remaining entries to $U(N_f -1)_D$ (the remaining $Q,\overline{Q}$ quarks).  

To include SM gauge fields, we will gauge some of the unbroken flavor symmetries of the theory.   In particular, we can weakly gauge a subgroup of $U(1)_D \times U(N_f -1)_D$, which we denote by
\be
U(1)_H \times SU(2)_F \times SU(3)_C \times U(1)_V \subset U(1)_D \times U(N_f -1)_D
\ee
with some malice aforethought.  The $SU(3)_C$ gauge bosons will be directly identified with the gluons of QCD, and weak $SU(2)_L$ will emerge in the far IR as a linear combination of $SU(2)_F$ and the dual magnetic gauge group.  Hypercharge will be identified with a linear combination of the $U(1)$'s, and the generator of $U(1)_H$ is
\be
H = \frac{1}{6} \, \diag(1, -2,-2, -1, -1, -1, \underbrace{1,\ldots, 1}_{N_f-6}).
\label{eq:Hgen}
\ee
Note that this generator commutes with the $SU(2)_F \times SU(3)_C$ generators, and the choice of $H$ is dictated by requiring no states with exotic hypercharges in the IR spectrum.\footnote{One viable deformation is to shift the final $N_f-3$ entries in the hypercharge generator by an integer. That theory would have heavy particles with exotic hypercharges, but they would still be able to decay.}  The $U(1)_H$ generator is, in general, not traceless so it is not orthogonal to $U(1)_V$.  This means that in the UV there will be some kinetic mixing between the two $U(1)$ gauge bosons, but this has little effect in the IR.

Gauge coupling unification is challenging to achieve in any composite theory, and our model is no exception. The presence of so much additional matter charged under the SM gauge groups imperils perturbative gauge coupling unification. Even the simplest generic model ($N_f = 7, \, N_c = 5$) leads to a Landau pole in $SU(3)_C$ somewhere between two and four orders of magnitude below the GUT scale, depending on the precise hierarchy of scales. This may be remedied in a variety of ways, including strong unification \cite{Kolda:1996ea} or a separate (asymptotically free) Seiberg dual for $SU(3)_C$.  Alternatively, Landau poles may be avoided altogether by using the special minimal embedding ($N_f = 6, \, N_c = 4$) presented in \App{app:64} (for which $SU(3)_C$ remains perturbative up to the GUT scale), or perhaps a chiral embedding along the lines of \Ref{Behbahani:2010wh}. Whatever the solution, the prospective $SU(3)_C$ Landau pole lies above all the dynamical scales of interest, rendering it consistent to treat $SU(3)_C$ as a weakly gauged flavor symmetry for the purposes of our model.

Apart from the Higgs sector, the remaining chiral SM matter fields will be elementary degrees of freedom.  From the perspective of the strong dynamics, the interesting SM operators are color-invariant bilinears of SM chiral fields of the form $\CO^{u}_{ij} \sim q_i u^c_j$ and $\CO^d_{ij} \sim q_i d^c_j, l_i e^c_j$ that we will choose to transform as $(\mathbf{1},\mathbf{2})_{-1/2}$ and $(\mathbf{1},\mathbf{2})_{1/2}$ under the gauged $SU(3)_C \times SU(2)_F \times U(1)_H$ flavor symmetries. Here, $i,j = 1,...,3$ are SM flavor indices, and these operators should be thought of as having canonical scaling dimension $\Delta_{\CO} = 2$.

We can add additional UV deformations which preserve the SM gauge symmetry, such as supersymmetric mass terms for the electric quarks
\begin{eqnarray}
W_e = m_0 P \overline P + m_I^J Q^I \overline Q_J ,
\end{eqnarray}
where $I,J = 1,...,N_f-1$ run over the $SU(N_f -1)$ flavor indices. 
The simplest choice of mass matrix that preserves the SM gauge symmetries and leads to the desired pattern of IR dynamics is
\be
m_I^J = \text{diag}(m_1, m_1,  \underbrace{m_2,\ldots,m_2}_{N_f-3}).
\ee
This leaves a diagonal $U(1) \times SU(2)_F \times  SU(N_f - 3) \times U(1)_V$ subgroup of the full global symmetries unbroken, where $SU(3)_C \subset SU(N_f - 3)$.\footnote{We may alternately choose masses $m_I^J$ that preserve only $SU(2)_F$ and $SU(3)_C \subset SU(N_f-3)$, provided that the desired mass hierarchies are preserved.}  For the sake of calculability, we will be interested in the case $m_i \ll \Lambda$, and a hierarchy $m_1 \gtrsim m_2 \gg m_0$.  With these mass deformations, this theory is known to possess metastable nonsupersymmetric vacua \cite{Intriligator:2006dd}.

Inspired by \Ref{Green:2010ww}, we will also add vector-like pairs of spectator fields $S,\overline S$ that are singlets under the strong dynamics but transform as complete multiplets under the $SU(N_f -1)_D$ global symmetry.  These states in fact have well-defined $SU(N_f -1)_L \times SU(N_f -1)_R$ quantum numbers as shown in \Fig{fig:UVMoose}.  Under the explicit breaking $SU(N_f - 1)_D \to SU(2) \times SU(N_f - 3)$, these spectators decompose as, e.g., $S = (S_{2}, S_{N_f-3})$.  These spectator fields will serve two purposes: to decouple unwanted dual degrees of freedom and to connect SM chiral matter to the SQCD fields.  In particular, we may add marginal couplings between the spectator fields and electric quarks as well as couplings between the spectator fields and the SM operators $\CO^{u,d}_{ij}$, all consistent with the unbroken flavor symmetries of the theory:
\begin{eqnarray} \label{eqn:UVdef}
\delta W_e &=& \lambda \, S \overline P Q + \overline \lambda \,  \overline{S} P \overline{Q}  - y^u_{ij} \overline{S}_2 \CO^u_{ij} - y^d_{ij} S_2 \CO^d_{ij} .
\end{eqnarray}
Our approach shares the philosophy of bosonic technicolor \cite{Samuel:1990dq,Dine:1990jd}, in that $S_2, \overline S_2$ are elementary fields with the quantum numbers of the Higgs doublets $H_d, H_u$ that will be used to induce couplings to  composite fields in the IR.

Finally, we will be interested in a deformation that breaks an accidental $R$ symmetry in the IR. The simplest such operators are quartic single- and double-trace operators for the electric quarks of the form
\begin{equation} 
\delta W_e = \frac{c_Q}{2\Lambda_0} \tr (Q \overline{Q})^2 + \frac{\gamma c_Q}{2\Lambda_0} (\tr Q \overline{Q})^2 + \frac{c_N}{2\Lambda_0} (P \overline P)^2.
\label{eq:multitrace}
\end{equation}
which may be induced by integrating out a massive adjoint at the scale $\Lambda_0$; as we will see, the scale should be such that $ c_Q \Lambda^2 / \Lambda_0 \sim$ TeV. Such deformations were studied in detail in \Ref{Essig:2008kz}. 

\subsection{The Magnetic Theory}
\label{sec:magnetictheory}

Below the scale $\Lambda$, this theory possesses a weakly-coupled description in terms of an $SU(2)_M$ magnetic gauge group with $N_f$ flavors of fundamental and antifundamental magnetic quarks $q, \overline q$ and a magnetic meson $M$. We will assume, for simplicity, that the scales of the UV and IR theories match, so that no intermediate scale appears in the superpotential. 

Under the $U(1)_D \times U(N_f-1)_D$ flavor symmetry, these fields may be decomposed as
\begin{equation}
M = \left( \begin{array}{cccc}
N & \overline \Sigma   \\
\Sigma & \Phi 
   \end{array}  \right), \qquad q^T = \left( \begin{array}{c} H_u \\ \chi \end{array} \right), \qquad \overline{q} = \left(\begin{array}{c} H_d \\ \overline \chi \end{array} \right) .
\label{eq:decomp}
\end{equation}
Our notation reflects the fact that some of the magnetic quarks ($H_u$, $H_d$) will be identified as the SM Higgs superfields, and one of the magnetic mesons ($N$) will be identified with an NMSSM-like singlet field.  The relevant transformation properties of the fields in the magnetic theory are given in \Fig{fig:IRMoose} and in \Tab{tab:MagFieldcontent}. 

\FIGURE[t]{
\includegraphics[scale=0.7]{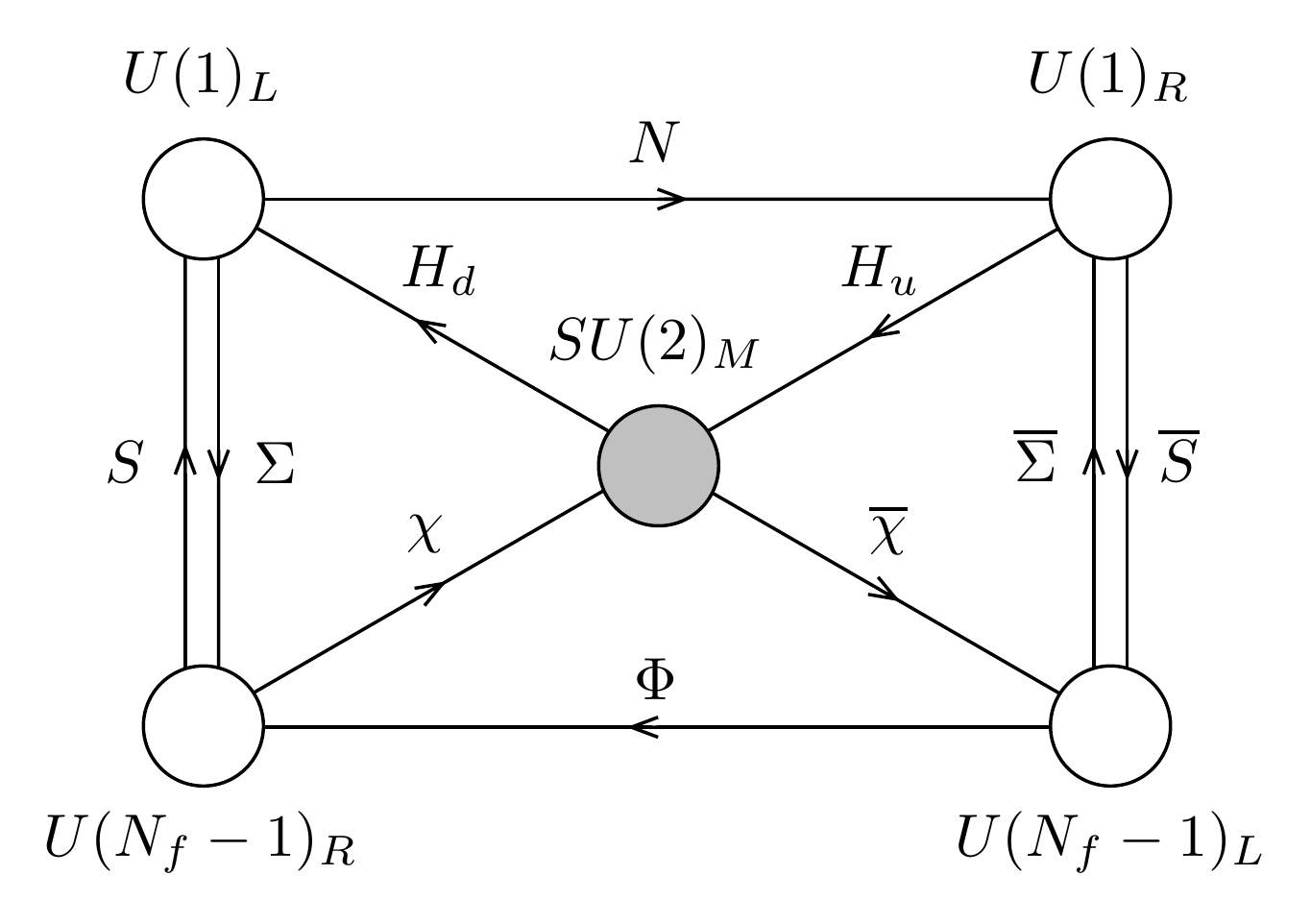}
\caption{The IR field content in moose notation.  The global symmetries are the same as in \Fig{fig:UVMoose}, but a dual magnetic gauge group $SU(2)_M$ appears below the confinement scale $\Lambda$.  The fields $H_u$, $H_d$, and $N$ will participate in an NMSSM-like Higgs sector, while the remaining dual fields will break SUSY as well as act as messengers.}
\label{fig:IRMoose}
}

  \TABLE[t]{
  \parbox{5in}{
  \begin{center}
\begin{tabular}{|c|c|cc|c|}
\hline
 & $SU(2)_M$ & $U(1)_D$ & $U(N_f-1)_D$ & $U(1)_V$  \\ \hline
$H_u$ & ${\bf 2}$ & $+1$ & ${\bf 1}$ & $+ 1/3$ \\
$H_d$ & ${\bf 2}$ & $-1$& ${\bf 1}$ & $-1/3$ \\
$\chi$ & ${\bf 2}$ & $0$ & $\bf \square$ & $+1/3$ \\
$\overline \chi$ & $ {\bf 2}$ & $0$& ${\bf \overline \square}$ & $-1/3$ \\ \hline
$N$ & ${\bf 1}$ & 0& ${\bf 1}$ & 0 \\ 
$\Sigma$ & {\bf 1} & $+1$& ${\bf \overline \square}$ & 0 \\
$\overline{\Sigma}$ & {\bf 1} & $-1$& $\bf \square$ & 0 \\
$\Phi$ & {\bf 1} &0 & {\bf Adj + 1} & 0 \\ \hline
$S$ & ${\bf 1}$ & $-1$& $\bf \square$ & $0$ \\
$\overline S$ & ${\bf 1}$ & $+1$ &  ${\bf \overline\square}$ & $0$ \\ \hline
\end{tabular}
\end{center}
}

\caption{The IR field content.  Unlike in \Fig{fig:IRMoose}, here we only give the quantum numbers under the diagonal $U(1)_D \times U(N_f-1)_D$ flavor symmetry, and have again identified $U(1)_V$, which is now magnetic quark number.   \label{tab:MagFieldcontent}}
}

The vectors and matrices in \Eq{eq:decomp} live in $SU(N_f)$ flavor space, with the second row and column carrying an $SU(N_f-1)_D$ index.  The fields $\Phi, \chi, \overline \chi$ decompose further under the explicit breaking $SU(N_f-1)_D \to SU(2)_F \times SU(N_f - 3)$ as 
\begin{equation}
\Phi = \left( \begin{array}{cccc}
Y & \overline Z \\
Z & X 
   \end{array}  \right), \qquad \chi^T = \left( \begin{array}{c} \sigma \\ \rho \end{array} \right), \qquad \overline \chi = \left(\begin{array}{c} \overline \sigma \\ \overline \rho \end{array} \right).
\label{eq:decomp2}
\end{equation}
This completes the notation for the IR degrees of freedom. 

The magnetic superpotential consists of magnetic Yukawa couplings dictated by duality, plus the appropriate mapping of the superpotential terms in the electric theory:
\begin{eqnarray}\label{eqn:wquark}
W_m &=& h N H_u H_d + h \chi_I \Phi^I_J \overline \chi^J  
+h H_u \overline \Sigma \overline \chi + h \chi \Sigma H_d \\ \nonumber
&& ~ -  h \mu_N^2 N - h (\mu^2)_I^J \Phi^I_J + \frac{1}{2}h^2 m_\Phi \Phi^I_J \Phi^J_I + 
\frac{1}{2} h^2 \gamma m_\Phi (\Phi^I_I)^2 + \frac{1}{2} h^2 m_N N^2 \\ \nonumber
&& ~ +  \lambda h \Lambda S \Sigma + \overline \lambda h \Lambda \overline{S} \overline{\Sigma} - y_{ij}^u \overline{S}_2 \CO_{ij}^u - y^d_{ij} S_2 \CO_{ij}^d.
\end{eqnarray}
The first line consists of Yukawa couplings dictated by Seiberg duality; the second line consists of source and mass terms arising from our UV mass deformations; and the third line arises from the spectator couplings introduced in \Eq{eqn:UVdef}.  The coupling $h$ is naturally $\mathcal{O}(1)$ and tracks the wavefunction renormalization of the meson fields. Up to incalculable factors of wavefunction renormalization (but retaining explicit powers of $h$), the parameters in the magnetic and electric theories are related by
\be
-h (\mu^2)_I^J = m_I^J \Lambda, \qquad -h \mu_N^2 = m_0 \Lambda, \qquad h^2 m_\Phi = c_Q \frac{\Lambda^2}{\Lambda_0},  \qquad h^2 m_N = c_N  \frac{\Lambda^2}{\Lambda_0}.
\ee

\FIGURE[t]{
\includegraphics[scale=0.7]{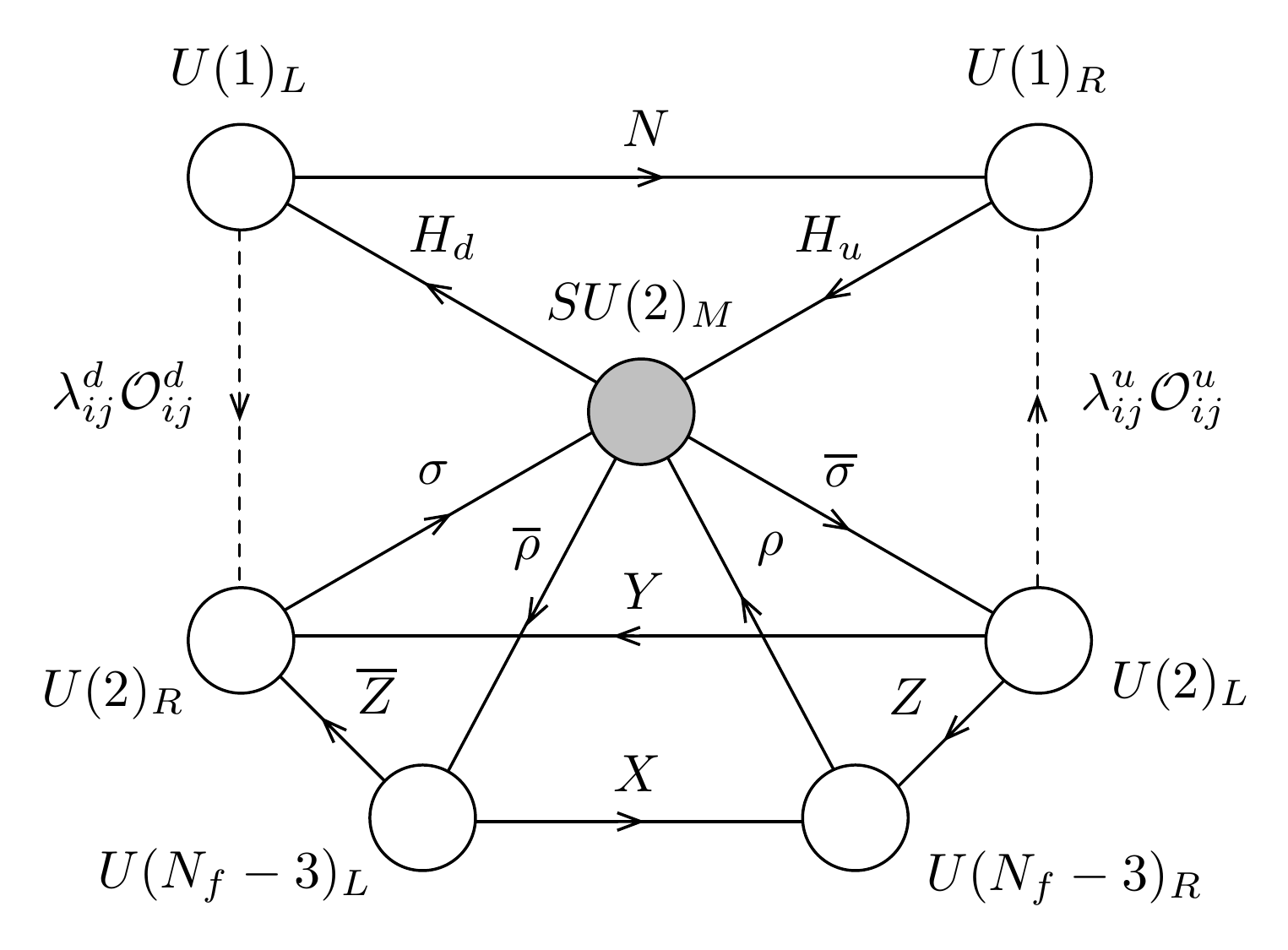}
\caption{The IR field content after integrating out the $S$, $\overline{S}$ spectator fields.  Here, we have made explicit the $U(2) \times U(N_f -3)$ subgroups of relevant $U(N_f-1)$.  Again, note that the labels $L$ and $R$ have been flipped in certain cases for clarity.  The dashed lines indicate chiral SM operators which have induced couplings to $H_u$, $H_d$ and $\sigma$, $\overline{\sigma}$.}
\label{fig:IRMooseNoSigma}
}

The UV coupling between spectators and electric quarks leads, in the IR, to a mass term between the spectators $S$, $\overline{S}$ and corresponding meson components $\Sigma$, $\overline{\Sigma}$.  Taking $\lambda = \overline \lambda$ for simplicity,  these fields may be integrated out at the scale $\lambda h \Lambda$, leading to the reduced field content in \Fig{fig:IRMooseNoSigma}.  Below the scale $\lambda h \Lambda$, the superpotential takes the form
\begin{eqnarray} \label{eqn:wmag}
W_m &=& h \chi_I \Phi^I_J \overline \chi^J   - h (\mu^2)_I^J \Phi^I_J + 
\frac{1}{2} h^2 m_\Phi \Phi^I_J \Phi^J_I + \frac{1}{2} h^2 \gamma m_\Phi (\Phi^I_I)^2\\ \nonumber
&& ~+  h N H_u H_d - h \mu_N^2 N + \frac{1}{2} h^2 m_N N^2  + \frac{1}{\lambda h \Lambda} y_{ij}^u H_u \overline \sigma \CO_{ij}^u + \frac{1}{\lambda h  \Lambda} y^d_{ij} H_d  \sigma \CO_{ij}^d .
\end{eqnarray}
The first line contains interactions that will lead to SUSY and $R$-symmetry breaking, while the second line contains interactions that will lead to EWSB and the generation of fermion masses.  

This construction is reminiscent of \Ref{Green:2010ww}, since after integrating out the heavy fields we have two nearly decoupled sectors, one containing $H_u$, $H_d$, and $N$, and one containing the rest of the SQCD fields. These two sectors only talk to each other through higher-dimensional operators suppressed by $\lambda h \Lambda$.  In what follows, we will neglect contributions from higher-dimensional  K\"ahler operators, which will provide $\mathcal{O}[(\mu_1 / \lambda h \Lambda)^2]$ corrections to the leading expressions.  At the end of the day, we will have to take $\mu_1 / \lambda h \Lambda$ to be close to one in order to have have a sufficiently large top Yukawa coupling, so these K\"ahler corrections will be parametrically (but not numerically) suppressed.

\section{SUSY Breaking and a Fat Higgs}
\label{sec:SUSYbreaking}

The magnetic superpotential in \Eq{eqn:wmag} leads to metastable SUSY-breaking vacua with spontaneously broken $R$-symmetry along the lines of \Ref{Intriligator:2006dd}.  In this section, we describe the dynamics at and immediately below the scale of SUSY breaking.  We will see how SUSY breaking leads to color-flavor locking, leading to heavy gauge boson masses and SM Yukawa couplings to the fat Higgs.  We will also see how SSM soft masses are generated via Higgsed gauge mediation.  A subsequent description of EWSB appears in \Sec{sec:EWSB}.

\subsection{Metastable SUSY Breaking and Color-Flavor Locking}

The fields $\{\Phi, \chi, \overline \chi \}$ comprise a sector breaking SUSY by the rank condition \cite{Intriligator:2006dd}. Specifically, the hierarchy $\mu_1 > \mu_2$ guarantees the existence of a metastable nonsupersymmetric vacuum in which $\sigma, \overline \sigma$ obtain vevs. There are not enough independent degrees of freedom to cancel the $F$-terms of $X$, defined in the decomposition of \Eq{eq:decomp}, so $|F_{X}| = |h \mu_2^2|$ and SUSY is broken with vacuum energy $V_0 = (N_f - 3) |h^2 \mu_2^4|$.  As discussed further in \Sec{subsec:susybreakingspectrum}, tree-level flat directions are all stabilized at one loop. The explicit $R$-symmetry breaking quartic deformations lead to a larger spontaneous $R$ breaking in which $X$ obtains a (small) nonzero vev.  This will eventually lead to gaugino masses proportional to $\langle X \rangle$. 

As shown in \Ref{Essig:2008kz}, the nonsupersymmetric vacuum lies at
\begin{eqnarray} 
\langle \sigma \rangle = \langle \overline \sigma \rangle &=& \mu_1 \delta^\alpha_a, \\ \nonumber
\langle \rho \rangle = \langle \overline \rho \rangle &=& 0, \\ \nonumber
\langle X \rangle &\approx& \frac{\mu_2^2 m_\Phi [1 + (N_f-3) \gamma]}{h b |\mu_2^2/\mu_1|^2} \delta_{c}^d, \\ \nonumber
\langle Y \rangle &=& 0,
\end{eqnarray} 
where $b = \frac{\log 4 - 1}{4 \pi^2}$ and $\gamma$ is the parameter of the $R$-symmetry-breaking deformation defined in \Eq{eq:multitrace}.  Here, $\alpha$ is an $SU(2)_M$ index, $a$ is an $SU(2)_F$ index, $c$ and $d$ are $SU(N_f-3)$ indices.  The vevs of $\sigma$ and $\overline{\sigma}$ break $SU(2)_M \times SU(2)_F \to SU(2)_L$, where $SU(2)_L$ is the SM electroweak symmetry.  This is an interesting example of magnetic color-flavor locking that will prove crucial in generating SM Yukawa couplings in \Sec{sec:farIR}.\footnote{The terminology ``color-flavor locking'' is taken from ordinary SQCD where $SU(2)_F$ is a global flavor symmetry.  Here, of course, $SU(2)_F$ is weakly gauged.}

This symmetry breaking pattern leads to three heavy $SU(2)$ gauge bosons of mass
\be
m_V^2 = 2 (g_M^2 + g_F^2) \mu_1^2
\ee
and three massless gauge bosons of $SU(2)_L$ with gauge coupling
\be
\label{eq:irEWcoupling}
\frac{1}{g^2} = \frac{1}{g_M^2} + \frac{1}{g_F^2}.
\ee 
The $\sigma$ and $\overline{\sigma}$ vevs also break $U(1)_V\times U(1)_H \to U(1)_Y$, where $U(1)_Y$ is SM hypercharge. We have normalized the generators in \Eq{eq:Vgen} and \Eq{eq:Hgen} such that
\be
Y=V+H,
\ee
and the fields $H_u, H_d$ have the usual hypercharge $Y=\pm 1/2$. The mass of the heavy $U(1)$ gauge bosons and low energy $U(1)_Y$ gauge coupling are analogous to the $SU(2)$ groups.  The vacuum alignment of $\vev{X}$ ensures that the diagonal $SU(N_f-3)$ flavor symmetry remains unbroken.  

\subsection{Spectrum of the SUSY-Breaking Sector}
\label{subsec:susybreakingspectrum}

We now consider the resulting mass spectrum in the SUSY-breaking sector $\{\Phi, \chi, \overline \chi \}$, with more details given in \App{app:soft}.  Most of these states get a mass at or a loop factor below the SUSY breaking scale, with the notable exception of the goldstino (eaten by the gravitino) and some exotic pseudo-modulini which will feature in the phenomenology described in \Sec{sec:pheno}.

The fermions $\psi_{\rho}, \psi_{Z}$ and $\psi_{\overline \rho}, \psi_{\overline Z}$, respectively pair up to obtain supersymmetric Dirac masses of order $h \mu_1$ from the vev of $\sigma$. The corresponding scalars combine into massive complex fields through various linear combinations of $\rho, \overline \rho^*, Z, \overline Z^*$, and most of these fields obtain masses of order $h \mu_1$ and splittings of order $h \mu_2$.   Of these, $4(N_f-3)$ real scalars obtain tree-level masses of order $h \sqrt{\mu_1^2 - \mu_2^2}$; had we not explicitly broken $SU(N_f-1)_D \to SU(N_f-3) \times SU(2)_F$ by $\mu_1 \neq \mu_2$, these would have been Nambu-Goldstone Bosons (NGBs) of {\it spontaneous} $SU(N_f-1)_D \to SU(N_f-3) \times SU(2)_F$ breaking. Properly speaking, the chiral superfields of the $\{\rho, Z \}$ sector are messengers of SUSY breaking, with $\CO(\sqrt{F_X} \sim h \mu_2)$ splittings between the fermions and scalars.

In the $\{Y, \sigma\}$ sector, fermions from $Y, \sigma + \overline \sigma$ pair up to form Dirac fermions with mass $h \mu_1$. The traceless part of the chiral superfield $\sigma - \overline \sigma$ contains the NGBs $\Im(\sigma' - \overline \sigma')$, which are eaten by the super-Higgs mechanism to give masses to the heavy gauge bosons of $SU(2)_M \times SU(2)_F \to SU(2)_L$; the corresponding real part obtains a mass of the same order as the heavy $SU(2)$ gauge bosons. (Here and henceforth, primes denote the traceless part of various fields.) These fields also obtain soft masses at one loop, which will lead to non-decoupling $D$-terms that raise the Higgs quartic coupling.  Of the trace part $\tr (\sigma - \overline \sigma)$, the chiral superfield $\Im \tr(\sigma - \overline \sigma)$ is an NGB associated with $U(1)_V \times U(1)_{H} \to U(1)_Y$ and is eaten by the super-Higgs mechanism.  Similarly, the chiral superfield $\Re \tr(\sigma - \overline \sigma)$ obtains a mass of the same order as the heavy $U(1)$ gauge bosons, as well as one-loop soft masses for the scalar components.

In the $X$ sector, the fermions $\psi_{X'}$ obtain masses from $R$-symmetry breaking of order $(N_f - 3) h^2 \gamma m_\Phi$, while the trace component $\psi_{\tr X}$ is the goldstino.  To get successful gaugino masses, $m_\Phi$ will turn out to be of order the TeV-scale, so these $\psi_{X'}$ fields can play a role in LHC physics.  The scalar components of $X$, on the other hand, are heavier.  The phase of the trace $\text{Arg}(\tr X)$ is an $R$-axion whose mass is of order $m_a^2 \sim h^3 m_\Phi \mu_2^2 / \langle X \rangle$, while the amplitude of the trace $|\tr X|$ and the traceless components $X'$ are pseudo-moduli that obtain masses at one loop via the Coleman-Weinberg (CW) potential \cite{Coleman:1973jx}, 
\be
V_{\CW} = \frac{1}{64 \pi^2} {\rm STr} M^4 \log \frac{M^2}{\Lambda^2}.
\ee
Here $M$ is the complete mass matrix as a function of $X', |\tr X|.$  In particular, near the origin of moduli space, this leads to positive masses for these pseudo-moduli of order 
\be
m_{\CW} \approx \frac{h}{4 \pi} \frac{h \mu_2^2}{\mu_1}.
\ee
Significantly, we find that there are no massless, charged fields arising in the SUSY-breaking sector, and all scalar fields obtain a positive mass-squared at tree level or one loop. 

There are, of course, a variety of supersymmetric vacua in addition to the nonsupersymmetric vacuum studied here. Explicit $R$-symmetry breaking leads to a supersymmetric minimum at $X \sim \mu^2_2/m_\Phi$, and transitions from the nonsupersymmetric vacuum to the supersymmetric one are exponentially suppressed by the small parameter $m_\Phi^2/b \mu_2^2$. Likewise, there is another set of supersymmetric vacua generated by irrelevant nonperturbative dynamics; transitions to these vacua are exponentially suppressed by $\mu_2^2/\Lambda^2$. The hierarchy $m_\Phi \ll \mu_{2} \lesssim \mu_1 \ll \Lambda$ ($m_i \ll \Lambda \ll \Lambda_0$ in the UV theory) therefore guarantees that the nonsupersymmetric vacuum is parametrically long-lived. Numerical analysis confirms that metastable vacua can easily be very long lived even in the presence of spectators and other superpotential deformations~\cite{Tamarit:2011ef}.  

Also note that many of the above states are stable at the level of the IR superpotential.  This is a generic feature of many SUSY breaking and mediation schemes, since they often involve large (unbroken) approximate global symmetries.  In \Sec{subsec:coloredexotics}, we will describe how the phenomenologically relevant states can be induced to decay.

\subsection{A Fat Higgs in the Far Infrared}
\label{sec:farIR}

Let us now turn to consider the effects of SUSY breaking on the $\{ N, H_u, H_d \}$ sector and elementary fields charged under the SM.  Below the scale of SUSY breaking, we can set the SUSY breaking fields to their vevs, and we are left with the fields $N, H_u, H_d$ and the superpotential
\begin{eqnarray}
\label{eqn:wIR}
W_{\rm IR} = h N H_u H_d - h \mu_N^2 N + \frac{1}{2} h^2 m_N N^2 + \frac{y_{ij}^u \mu_1}{\lambda h \Lambda} H_u\CO_{ij}^u + \frac{y_{ij}^d \mu_1}{\lambda h \Lambda} H_d \CO_{ij}^d.
\end{eqnarray}

Several comments are in order. 
\begin{itemize}
\item This has the same field content as the Higgs sector of the NMSSM, but instead of a trilinear term for the singlet, we have a linear source term.  This is the signature of a so-called ``fat'' Higgs \cite{Harnik:2003rs}.

\item The first two terms in \Eq{eqn:wIR} break electroweak symmetry in the supersymmetric limit. This is in contrast to the MSSM where electroweak symmetry can only be broken after SUSY is broken. The effect of the third term depends on the size of $m_N$ relative to $\mu_N$, as we will discuss in further detail in \Sec{sec:EWSB}.

\item   The last two terms are Yukawa couplings between the Higgs fields and SM fermions.  The magnetic color-flavor locking vevs for $\sigma, \overline \sigma$ have converted the previously irrelevant couplings involving $\CO^{u,d}$ into marginal Yukawa couplings between the SM operators and magnetic quarks $H_u, H_d$. 
\end{itemize}
In \Sec{sec:EWSB}, we will see that these features lead to successful EWSB.

In order to have a sufficiently large top Yukawa coupling, we require $y_{33}^u \mu_1 \simeq \lambda h \Lambda$.  As discussed at the end of \Sec{sec:magnetictheory}, this implies that there will be large K\"ahler corrections, but the basic vacuum structure and qualitative spectrum are unchanged.  This theory does not automatically explain the hierarchy of SM flavor, but suitable flavor textures may be generated in the UV involving any combination of the couplings between SM fields, electric quarks, and spectator fields.  In particular, we have taken $\lambda = \overline{\lambda}$ for simplicity, but if desired, an up/down hierarchy could be generated by splitting the spectator masses.

\subsection{SSM Soft Spectrum}
\label{sec:MSSMsoft}

We finally turn to the impact of SUSY breaking on the SSM degrees of freedom.   SUSY breaking in the $\{\Phi, \chi, \overline \chi \}$ sector leads to soft terms for both the magnetic Higgs sector degrees of freedom $ N, H_u, H_d$ and the elementary SM fields. The primary source of soft masses is merely gauge mediation, which gives positive masses to all relevant scalar degrees of freedom and ensures stability of the vacuum.  In particular, the modes $\rho,\overline{\rho}$ get tree-level SUSY breaking mass splittings and mix with the $Z, \overline Z$ modes once $\sigma,\overline \sigma$ obtain a vev; the $\{ \rho, Z\}$ sector therefore constitute messengers with messenger mass $M \simeq h \mu_1$ and SUSY breaking scale $F \simeq h \mu_2^2$.   Gaugino masses arise as a result of $R$-symmetry breaking and are parametrically different from the scalar soft masses. 

Note that the particular gauge-mediated spectrum is somewhat unusual.  Since both $SU(2)_F \times SU(2)_M \to SU(2)_L$ and $U(1)_H \times U(1)_V \to U(1)_Y$ at the scale of SUSY breaking, the soft masses at the SUSY-breaking scale are those of Higgsed gauge mediation \cite{Gorbatov:2008qa}. Moreover, as the scale of Higgsing is well above the scale of gaugino masses, there is only one light gaugino from each of $SU(2)_F \times SU(2)_M \to SU(2)_L$  and  $U(1)_H \times U(1)_V \to U(1)_Y$, each of which behave as superpartners of the massless  $SU(2)_L \times U(1)_Y$ SM gauge bosons to excellent approximation.  Hence, the only gauginos appearing in renormalization group (RG) evolution of soft masses down to the weak scale are the conventional gluino, wino, and bino.  

The contributions to soft masses from gauge mediation are as follows.
\begin{itemize}
\item \textbf{Gauginos}: The gauginos obtain masses proportional to $R$-symmetry breaking, of order 
\begin{equation}
m_{\lambda_a} \simeq g_a^2 [1 + (N_f-3) \gamma] m_\Phi \left(\frac{\mu_2}{\mu_1} \right)^4 ,
\end{equation}
where $a$ labels the SM gauge groups.  This means that the usual GUT relation for the SM gauginos holds approximately, up to subleading corrections due to the complicated messenger sector.  Note that while the dominant contribution would be expected at $g_a^2 m_\Phi$, the additional suppression comes from the fact that gaugino masses vanish at leading order in $F/M^2$. Conversely, the loop factor is cancelled by the inverse loop factor coming from spontaneous $R$-symmetry breaking by the $X$ vev. 

\item \textbf{Sfermions}: The sfermions obtain masses unsuppressed by $R$-symmetry breaking. The masses-squared may be written as a sum of two contributions: one conventional contribution coming from the massless gauge bosons, and one additional contribution coming from the massive gauge bosons, suppressed relative to the first contribution by both $(g_F/g_M)^4$ and additional numerical coefficients $a_1,a_2$.\footnote{For details of the calculation and the precise form of the numerical suppression, see \Ref{Gorbatov:2008qa}.  In our case, $a_1,a_2 \sim \mathcal{O}(0.5)$.} Hence 
\begin{eqnarray} \nonumber
m_{\tilde f}^2 &\simeq&  \left[ C_3^r \left( \frac{\alpha_3}{4 \pi} \right)^2 + C_2^r \left(1 + a_1 \frac{g_F^4}{g_M^4} \right) \left( \frac{\alpha_2}{4 \pi} \right)^2 \right. \\
&& \left. ~ + \frac{3}{5} Y^2  \left(1 + a_1 \frac{g_H^4}{g_V^4}  + 2 a_2 \frac{g_H^2}{g_V^2} \right) \left( \frac{\alpha_1}{4 \pi} \right)^2  \right] \left( \frac{\mu_2^2}{\mu_1} \right)^2 \, ,
\label{eqn:MatterSoft}
\end{eqnarray}
where the $C_a^r$ are the appropriate quadratic Casimirs of $SU(3)_C$ and $SU(2)_L$ for the representation $r$, and $Y$ is its hypercharge.

\item \textbf{Higgses}: The Higgses obtain their soft masses much in the manner of the other sfermions, though the additional contributions to their masses from heavy gauge bosons are {\it enhanced} by the ratio  $(g_M/g_F)^4$. The parametric differences between the Higgs and sfermion masses arise because the sfermions are charged under $SU(2)_F$, while the Higgses are charged under $SU(2)_M$. Hence 
\begin{eqnarray} \nonumber
m_{H_u, H_d}^2 &\simeq& \left[ C_2^r \left(1 + a_1 \frac{g_M^4}{g_F^4} \right) \left( \frac{\alpha_2}{4 \pi} \right)^2\right. \\
&& \left. ~ + \frac{3}{5} Y^2  \left(1 + a_1 \frac{g_V^4}{g_H^4}  + 2 a_2 \frac{g_V^2}{g_H^2} \right) \left( \frac{\alpha_1}{4 \pi} \right)^2  \right] \left( \frac{\mu_2^2}{\mu_1} \right)^2.
\label{eqn:HiggsSoft}
\end{eqnarray}
The mass for $H_u$ is also significantly reduced by RG evolution to the weak scale due to the size of the top Yukawa coupling. 

\item \textbf{Singlet}: The singlet does not get a soft mass from gauge mediation because it is neutral under all the gauge symmetries. There is a small tachyonic soft mass of order $m_S^2 \sim - \frac{h^4}{(16 \pi^2)^2} \frac{\mu_2^4}{\lambda^2 \Lambda^2}$ generated by the two-loop CW potential \cite{Giveon:2008wp, Giveon:2008ne}. Perhaps more importantly, the soft masses for the Higgses feed into the RG evolution for the singlet soft mass and drive the soft mass more negative. Hence the singlet soft mass is typically of order
\begin{equation}
m_S^2 \simeq -4 \frac{h^2}{16\pi^2} m_{H_d}^2 \log(\mu_1^2/m_{H_d}^2).
\end{equation}
\end{itemize}
This completes the contributions from Higgsed gauge mediation.  Note that $A$- and $B$-terms are small and generated mostly by radiative effects, but they will be important in the discussion of \Sec{sec:EWSBmssm}.

Finally, there is an additional contribution to the soft masses of scalars charged under $SU(2)_L \times U(1)_Y$ coming from $D$-terms after inserting the vev $\langle \sigma \rangle$.  For sfermions, this additional contribution is of the form
\begin{equation}
\delta m_{\tilde f}^2 = \left(C_2^r \frac{g_F^2}{g_M^2} \frac{\alpha_2}{2 \pi} + \frac{3}{5} Y^2 \frac{g_V^2}{g_H^2} \frac{\alpha_Y}{2 \pi} \right)  m_{\CW}^2 + \CO(m_{\CW}^2 / m_V^2),
\end{equation}
where $m_{\CW}^2 \sim \frac{h^2}{8 \pi^2} \frac{h^2 \mu_2^4}{\mu_1^2}$ is the one-loop CW soft mass of the scalar $\delta \sigma_- = \frac{1}{\sqrt{2}} (\delta \sigma - \delta \overline \sigma)$, and additional corrections are suppressed by the smallness of this soft mass relative to the scale of the heavy gauge bosons.  For the Higgses, this additional contribution is of the form 
\begin{equation}
\delta m_{H_u, H_d}^2 = \left(C_2^r \frac{g_M^2}{g_F^2} \frac{\alpha_2}{2 \pi} + \frac{3}{5} Y^2 \frac{g_H^2}{g_V^2} \frac{\alpha_Y}{2 \pi} \right)  m_{\CW}^2 + \CO(m_{\CW}^2 / m_V^2).
\end{equation}
This is the well-known radiative correction to the Higgs soft masses that arises in theories with non-decoupling $D$-terms~\cite{Batra:2003nj, Maloney:2004rc}, whose effect on the Higgs we will discuss below. In order to avoid significant fine-tuning of the Higgs mass, it is necessary for the CW soft masses to be below $\sim 10$ TeV.

\section{Electroweak Symmetry Breaking}
\label{sec:EWSB}

We now turn to the dynamics of the Higgs sector arising from \Eq{eqn:wIR} and the associated soft masses.   Below the scale of SUSY breaking, the remaining IR dynamics drives EWSB.  The superpotential for the Higgs fields in \Eq{eqn:wIR} leads to a scalar potential 
\be
V_W = h^2 |H_d H_u - \mu_N^2 + h m_N N|^2 +h^2 |N|^2 \left(|H_u|^2 + |H_d|^2 \right).
\label{eqn:higgsPot}
\ee
As discussed in \Sec{sec:MSSMsoft}, the relevant soft SUSY breaking terms in the scalar potential are
\be
V_{\rm soft} = m_{H_u}^2 |H_u|^2 + m_{H_u}^2 |H_d|^2 + m_S^2 |N|^2,
\label{eqn:Vsoft}
\ee
with $m_{H_i}^2 \sim \frac{\alpha_M^2}{16 \pi^2} \frac{\mu_2^4}{\mu_1^2}$  and $m_S^2 \sim -\frac{h^2}{2\pi^2}m_{H_i}^2$, where $\alpha_M$ is the magnetic gauge group structure constant.  There is also a $D$-term potential to be discussed in \Sec{sec:quartic}.  

The actual pattern of EWSB depends sensitively on the relation between $m_N$ and $\mu_N$.  In the limit $m_N \ll \mu_N$, EWSB occurs in the supersymmetric limit, as discussed further in \Sec{sec:EWSBsusy}. In this case, it is not necessary for $m_{H_u}^2$ to run negative, and EWSB is driven largely by superpotential terms.  In contrast, in the limit $m_N \gg \mu_N$ we may integrate out $N$; we then recover a version of the MSSM with irrelevant operators, and electroweak symmetry may only be broken {\it nonsupersymmetrically}, as discussed further in \Sec{sec:EWSBmssm}.  This requires $m_{H_u}^2$ to run negative.  In the intermediate case $m_N \simeq \mu_N$,  the resulting vacuum typically preserves electroweak symmetry, so we will therefore focus on the dynamics in the hierarchical limits $m_N \ll \mu_N$ and $m_N \gg \mu_N$.  Significantly, both limits share various features arising from compositeness that raise the physical Higgs mass relative to the MSSM, primarily by enhancing the Higgs quartic coupling.

\subsection{A Large Higgs Quartic}
\label{sec:quartic}

The mass of the lightest Higgs scalar is controlled by the quartic, and in the MSSM the quartic comes only from $D$-terms and constrains the tree-level mass of the lightest Higgs boson to be lighter than the $Z$ boson.  Raising the mass by radiative corrections so that it is above the LEP bounds requires heavy stops (and/or large $A$-terms) and introduces a little hierarchy problem.  

Increasing the tree-level mass of the lightest Higgs scalar ameliorates the little hierarchy problem, and thus motivates many NMSSM constructions and fat Higgs models.  These models have an additional contribution to the quartic coming from the coupling to the singlet shown in \Eq{eqn:higgsPot}, which goes like $h^2$.  We emphasize that in our construction, the superpotential Yukawa coupling $h$ may be naturally quite large at the weak scale.  In conventional versions of the NMSSM, the size of this Yukawa coupling in the IR is limited by the desire to avoid a Landau pole at low scales.  However, in theory at hand, duality provides a natural and well-behaved UV completion at the scale $\Lambda$ that allows us to evade limits coming from perturbativity. 

Beyond the large $F$-term quartic, an additional contribution which does not appear in other fat Higgs models is a consequence of the mixing between $SU(2)_M$ and $SU(2)_F$ induced by SUSY breaking.  Loops of heavy $SU(2)$ gauge bosons give rise to a non-decoupling correction to the Higgs quartic $D$-term~\cite{Batra:2003nj, Maloney:2004rc, Craig:2011yk}.  The $D$-term in this theory may be readily computed by integrating out the fluctuation $\delta \sigma_- = \frac{1}{\sqrt{2}} \Im (\delta \sigma - \delta \overline \sigma)$ at tree level. This field obtains a supersymmetric mass-squared $2(g_M^2 + g_F^2) \mu_1^2$ from the super-Higgs mechanism, as well as a nonsupersymmetric mass $m_{\CW}^2 \sim \frac{h^2}{8 \pi^2} \frac{h^2 \mu_2^4}{\mu_1^2}$ at one loop. Integrating out $\delta \sigma_-$, the $D$-term for the diagonal electroweak $SU(2)_L$ exhibits a non-decoupling correction to the Higgs quartic of the form
\begin{equation}\label{eqn:dterm}
V_D = \frac{g_M^2 g_F^2}{8 (g_M^2 + g_F^2)} \left( 1 + \frac{g_M^2}{g_F^2} \frac{2 m_{\CW}^2}{2(g_M^2 + g_F^2) \mu_1^2 + 2 m_{\CW}^2} \right) \left | H_u^\dag \sigma^a H_u - H_d^\dag \sigma^a H_d \right |^2,
\end{equation}
where $m_{\CW}$ is the CW soft mass of $\delta \sigma_-$, $g_M$ is the magnetic $SU(2)$ gauge coupling, and $g_F$ is the coupling of the gauged $SU(2)$ flavor symmetry in the SUSY-breaking sector.  

We recognize $\frac{g_M^2 g_F^2}{g_M^2 + g_F^2}$ as simply being the IR $SU(2)_L$ gauge coupling squared $g^2$ from \Eq{eq:irEWcoupling}, so the overall scaling of the $D$-term is that same as for $SU(2)_L$ alone.  But only a few decades of energy lie between $\Lambda$ and the scale of $SU(2)_M$ breaking, meaning that $g_M$ is naturally quite large and the correction to the Higgs quartic is parametrically enhanced by an amount $g_M^2/g_F^2$.  Of course, the $m_{\CW}$ soft mass is suppressed relative to $2(g_M^2 + g_F^2) \mu_1^2$, but in generic regions of parameter space, these competing effects still give rise to an $\mathcal{O}(1)$ overall correction.  An identical contribution comes from the hypercharge $D$-term, as the scale of Higgsing is the same, though in this case the correction to the quartic is proportional to the ratio $g_V^2/g_H^2$, which is not necessarily large.

This $D$-term quartic correction raises the tree-level prediction for the Higgs mass by an amount 
\begin{equation}
\delta m_h^2 = \frac{g^2 \Delta + g'^2 \Delta'}{2} v^2 \cos^2(2 \beta),
\end{equation}
where $\Delta$ is the second term in parentheses in \Eq{eqn:dterm}, $\Delta'$ is the corresponding (subdominant) hypercharge term, and $\tan\beta$ is the usual ratio of the vevs of the Higgses.  For moderate values of $\tan \beta$, this provides a correction to the Higgs mass ranging between a few GeV (for $h=1$) to a few tens of GeV (for $h=2$). 

Thus, the magnetic nature of the Higgs enhances the quartic in two complementary ways, as we will see in more detail below.  In the supersymmetric NMSSM-like limit, we have $\tan \beta \simeq 1$, in which case the $D$-term corrections to the Higgs quartic are suppressed.  However, in this limit the $F$-term contributions to the Higgs quartic are maximal.  In contrast, in the nonsupersymmetric MSSM-like limit, the $F$-term contribution is diminished, but $\tan \beta \gg 1$ so that the $D$-term corrections are significant.  Whatever the parametric limit, a fat Higgs with a magnetic personality significantly outweighs its MSSM counterpart.

\subsection{EWSB in the Supersymmetric Limit}
\label{sec:EWSBsusy}

In the limit $\mu_N \gg m_N$, EWSB may occur supersymmetrically.  In the supersymmetric limit, the superpotential of \Eq{eqn:wIR} has two minima, only one of which breaks electroweak symmetry. Loosely speaking, these minima correspond to $\langle H_{u,d} \rangle \sim \mu_N, \langle N \rangle \sim 0$ and $\langle H_{u,d} \rangle \sim 0, \langle N \rangle \sim \mu_N^2 / h m_N$, respectively.  Once we include the nonsupersymmetric soft corrections from \Eq{eqn:Vsoft}, the electroweak-preserving vacuum is destabilized provided $m_N$ is not too large. 

While the analytic form of the EWSB vacuum is difficult to compute, we can understand the parametric behavior by making a few simplifying assumptions.  In particular, consider the case $m_{H_u}^2 \simeq m_{H_d}^2$.  This is equivalent to $\tan\beta \simeq 1$ because in this limit there is an exchange symmetry between the two Higgs doublets in the potential.  This holds to good approximation at the scale of SUSY breaking, where we have $m_{H_u}^2= m_{H_d}^2$, but RG running from the top Yukawa coupling will reduce $m_{H_u}^2$ relative to $m_{H_d}^2$ and raise $\tan\beta$ above 1.  This effect, however, is often expected to be small for two reasons.  First, the scale of SUSY breaking is quite low so there are only a few decades of running.  Second, the Higgs soft masses are larger than in usual gauge mediation models because of the contribution from loops of the heavy $SU(2)$ bosons parameterized by $a_1$ in \Eq{eqn:HiggsSoft}, so the running effects from Yukawa couplings can be relatively small.\footnote{Of course, there are regions of parameter space where both of these considerations are vitiated, as we will see in \Sec{sec:EWSBmssm}.} 

For $\tan\beta \simeq 1$, we can ignore the potential coming from the $D$-terms even though it is parametrically larger than in the MSSM. We can then express the EWSB parameters $v^2 = \langle H_u\rangle^2 + \langle H_d\rangle^2 \simeq (175 \; {\rm GeV})^2$ and $\tan\beta$ simply in terms of the superpotential parameters and soft masses:
\begin{eqnarray}
v^2 &=& \left( \mu_N^2 - \frac{m_{H_d} m_{H_u} }{h^2}\right)
           \left( \frac{m_{H_d}^2 + m_{H_u}^2}{m_{H_d} \, m_{H_u}} \right), \nonumber\\
\tan\beta &=& \frac{m_{H_d}}{m_{H_u}}, \nonumber\\
\langle N \rangle &=& \frac{m_N \, m_{H_u} \, m_{H_d}}{ h^2 v^2  + m_S^2},
\end{eqnarray}
where $m_S$ is the (tachyonic) soft mass of the singlet as in \Eq{eqn:Vsoft}, and we have ignored terms $O(m_N^2/\mu_N^2)$ and higher.  The computation of the vacuum structure is quite similar to~\Ref{Harnik:2003rs}.  From these equations we see that EWSB is driven by the supersymmetric parameter $\mu_N$ and thus occurs even in the supersymmetric limit. 
Viable EWSB happens provided the Higgs soft masses are smaller than $\mu_N$, and there is a reasonable (though small) effective $\mu$ term which is proportional to $m_N$.  

In the approximation $m_{H_u}^2= m_{H_d}^2 \equiv m_{H}^2$, there is a SM-like Higgs $h^0=(H^0_u + H^0_d)/\sqrt{2}$ whose mass is 
\be
m^2_{h^0} = h^2 v^2.
\ee
As advertized, while the MSSM $D$-term contribution vanishes when $\tan\beta=1$, the SM-like Higgs gets a large mass from the coupling to the singlet.   

As in any theory of two Higgs doublets, there is also a pseudoscalar Higgs $A^0$, a heavier neutral scalar Higgs $H^0$, and the charged scalar Higgses $H^{\pm}$.  In this limit, their masses are:
\begin{eqnarray}
m^2_{A^0} &=& 2 m_H^2 + h^2 v^2, \nonumber\\
m^2_{H^0} &=&2 m_H^2 + \frac{g^2(1+\Delta)+g'^2(1+\Delta')}{2}v^2,\\
m^2_{H^\pm} &=& 2 m_H^2 + \frac{1}{2}g^2(1+\Delta)v^2. \nonumber
\end{eqnarray}
This predicts a fixed ordering between the states, in which the pseudoscalar will be heaviest, followed by the heavy scalar, with the charged Higgses being lightest.  This is unlike the MSSM where the pseudoscalar is always lighter than the charged states.  Finally, there is a singlet scalar and pseudoscalar from the $N$ multiplet, which are degenerate and do not mix with the Higgs in the $m_N \to 0$ limit. Their masses are both given by 
\be
m^2_{s^0} = h^2 v^2 + m_S^2.
\ee
These states will typically be lighter than all the Higgs states because $m_S^2$, the soft mass for $N$, is tachyonic.

As in the NMSSM, the neutralino sector now has five states with the singlino mixing with the usual gauginos and higgsinos.  In the limit where $m_{\lambda_a} \gg hv \gg h \langle N \rangle$, the gauginos are roughly mass eigenstates with masses $m_{\lambda_a}$ as described in \Sec{sec:MSSMsoft}. The higgsinos and singlino have large mixing between them, with two getting mass $\sim h v$ and the third getting mass $g^2 v^2/m_{\lambda_2}$, which can be quite light.  
The regime in which $m_{\lambda_a} \sim hv$ is also quite interesting, and typically leads to a viable spectrum of neutralinos as well. While it is difficult to come up with analytic solutions to the eigenvalue equation in this regime, numerical study of the parameter space reveals that the lightest neutralino is often mostly bino, with mass between about 10 and 100 GeV.

The chargino mass matrix is the same as in the MSSM with the $\mu$ term coming from $h \langle N \rangle$. The chargino which is mostly wino will have mass $\sim m_{\lambda_2}$ while the mostly-higgsino chargino will have mass $\sim h \langle N \rangle  - g^2 v^2/m_{\lambda_2}$, so it is expected to lie not too far above the LEP bound for charged states.  Note that the phase of $h \langle N \rangle$ and $m_{\lambda_2}$ must be aligned or anti-aligned to avoid introducing new sources of $CP$ violation; if they are anti-aligned this raises the mass of the lightest chargino. 

The size of the magnetic gauge coupling plays an important role in the Higgs soft masses.  For a low duality scale, $g_M$ is large and the Higgs mass parameters are larger than other $SU(2)$ charged scalars. In this region, the RG running has a smaller effect and the spectrum looks quite similar to the one discussed so far in this section.  On the other hand, $g_M$ could be smaller which makes the spectrum more MSSM-like.  In \Fig{fig:spectrum} we show an example of each kind of spectrum. From the figure we see that small changes in $g_M$ can have a large effect on the Higgs soft masses which affects the entire scalar spectrum. 

\FIGURE[t]{
\includegraphics[scale=.55]{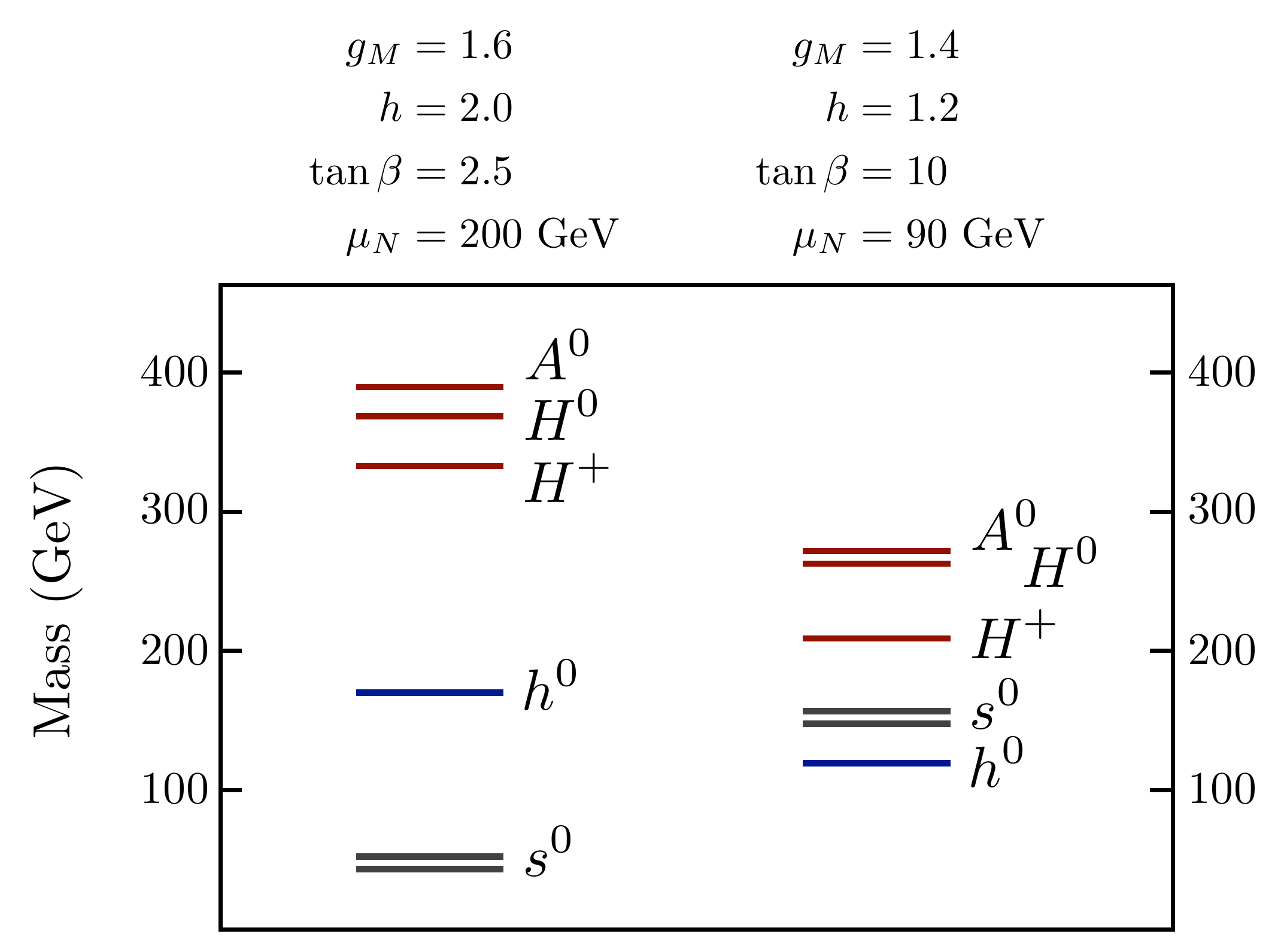}
\caption{Example scalar spectrum in the approximation of $m_N=0$ where the singlet states do not mix with the Higgses. Note that the value of $h$ is given at the scale $\mu_1=100$ TeV and will be decreased at the weak scale by RG running.   For the left spectrum, both $m_{H_u}^2$ and $m_{H_d}^2$ are positive at the weak scale, while for the right spectrum $m_{H_u}^2$ runs negative.}
\label{fig:spectrum}
}

The two spectra presented both have the SM-like Higgs with tree-level mass between 115 and 200 GeV, so they are allowed by precision electroweak measurements.  There is plenty of parameter space, however, where the SM-like Higgs can be heavier. The precision electroweak analysis in this case is entirely analogous to~\Ref{Harnik:2003rs}, which found such heavier Higgses to be compatible with precision electroweak constraints.

\subsection{EWSB in the Nonsupersymmetric Limit}
\label{sec:EWSBmssm}

Thus far, we have focused on the case for which EWSB proceeds supersymmetrically.  However, there is also a limit of the theory in which electroweak symmetry is broken nonsupersymmetrically much as in the MSSM, albeit with additional corrections that raise the tree-level prediction for the lightest neutral Higgs mass.

In the limit $m_N \gg \mu_N$, we may integrate out the singlet $N$ supersymmetrically at the scale $m_N$. This leaves us with an MSSM-like Higgs sector with additional corrections coming from irrelevant operators suppressed by $m_N$. In this limit, the Higgs sector superpotential below $m_N$ is
\be
\label{eqn:mssmlim}
W = \frac{\mu_N^2}{m_N} H_u H_d - \frac{1}{2 m_N} (H_u H_d)^2\,.
\ee
Thus, we find a conventional supersymmetric $\mu$ term of order $\mu_H = \mu_N^2 / m_N$, plus an additional quartic superpotential correction.  The physics of EWSB is simply that of the MSSM with certain irrelevant operators \cite{Dine:2007xi}. Successful EWSB then requires both that the combination $m_{H_u}^2 + \mu_H^2$ runs negative above the weak scale and that an adequate $B \mu$ term is generated. 

Radiative EWSB is typically easy to achieve in the MSSM even for the lowest-scale models of gauge mediation due to the significant RG effects of the top Yukawa. However, as discussed in \Sec{sec:EWSBsusy}, there is an enhancement of the Higgs soft masses  from the $SU(2)_M$ gauge coupling seen in \Eq{eqn:HiggsSoft}. This raises the UV values of the Higgs soft masses significantly above those of other electroweak-charged states, and it is far from clear that the soft masses will run negative. 

Interestingly, there are additional contributions in our construction that actually favor radiative EWSB. Above the scale $m_N$, the field $N$ runs in loops that renormalize the Higgs soft masses. The largeness of the $h$ Yukawa means that these contributions are as effective as the top Yukawa in driving the Higgs soft masses negative.  In the limit where we retain only $y_t$ and $h$ among the various Yukawa couplings, the one-loop RG equation for $m_{H_u}^2$ \emph{above the scale $m_N$} is 
\bea
16 \pi^2 \frac{d}{dt} m_{H_u}^2 = 3 X_t + X_N + \frac{3}{5} g_1^2 {\rm Tr}[Y_j m_{\phi_j}^2] - \frac{6}{5} g_1^2 m_{\lambda_1}^2 - 6 g_2^2 m_{\lambda_2}^2,
\eea
where $X_t = 2 |y_t|^2 (m_{\tilde Q_3}^2 + m_{H_u}^2+ m_{\tilde{\overline u}_3}^2) + 2 |a_t|^2$ and $X_N = 2 |h|^2  (m_{H_d}^2 + m_{H_u}^2 + m_N^2) + 2 |a_N|^2$. Here $a_t, a_N$ are the $A$-terms corresponding to the top and singlet Yukawa couplings, respectively; these are loop-suppressed relative to the other soft masses and typically negligible. Below the scale $m_N$, of course, we should integrate out $N$ and the RG equation is simply that of the MSSM. But for $m_N \lesssim 10$ TeV, the additional contribution from $h$ is often sufficient to drive $m_{H_u}^2$ negative provided adequate RG time. A representative illustration of the range of UV soft masses and messenger scales for which radiative EWSB occurs is shown in \Fig{fig:radewsb}. For lower values of the messenger scale ($\mu_1 \lesssim 10^3$ TeV), successful radiative EWSB requires significantly smaller values of $g_M$ (equivalently, higher values of $\Lambda$) in order to reduce the boundary value of $m_{H_u}$. It is important to note that the messenger scale $\mu_1$ may not be made arbitrarily large, lest the gaugino masses be too small.

\FIGURE[t]{
\includegraphics[scale=1]{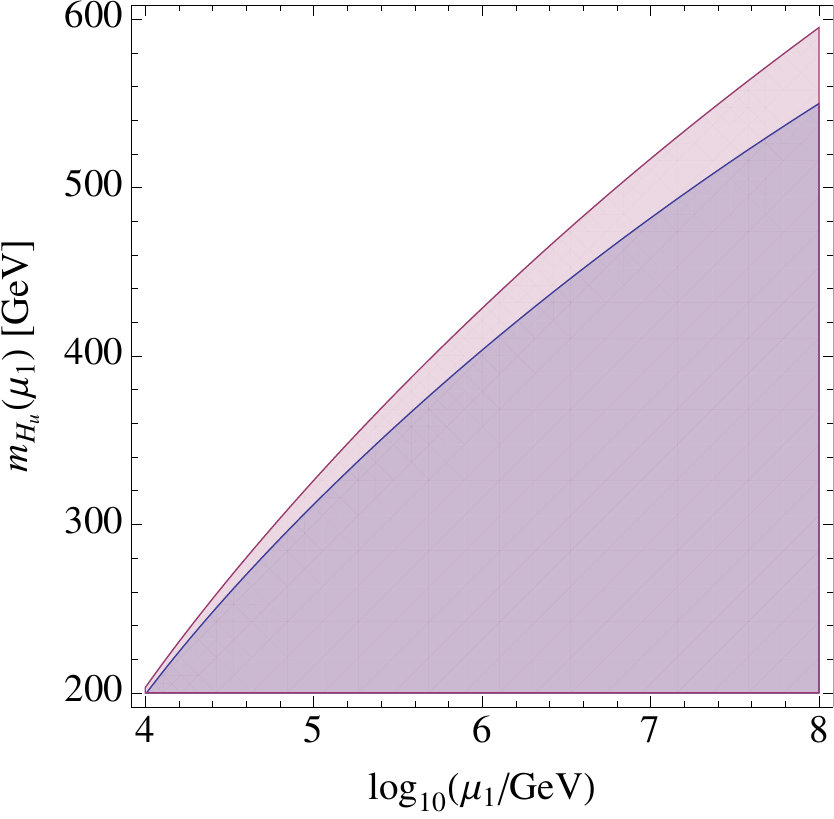}
\caption{Range of UV Higgs soft mass $m_{H_u}$ and messenger scale $\mu_1$ for which radiative EWSB is possible ($m_{H_u}^2 + \mu_H^2 < 0$) in the MSSM-like limit. The darker shaded region corresponds to $h(\mu_1)=1$, while the lighter region corresponds to $h(\mu_1)=2$. We have taken a representative soft spectrum and $\mu_N = 500$ GeV, $m_N = 2$ TeV. }
\label{fig:radewsb}
}

Since the effective $\mu$ term is generated supersymmetrically and direct gauge mediation yields $B \mu \sim 0$ at the scale of SUSY breaking, $B \mu$ is generated radiatively during RG evolution to the IR. Above the scale $m_N$, $\beta_{B \mu} \sim 0$ at two loops; $B \mu$ is generated radiatively only below $m_N$ and comes out around
\be
B \mu \sim - \mu_H \left( \frac{3 \alpha_2}{2 \pi} m_{\lambda_2} \log \frac{m_N}{m_{\lambda_2}} + \frac{3 \alpha_1}{10 \pi} m_{\lambda_1} \log \frac{m_N}{m_{\lambda_1}} \right).
\ee
(The actual value is somewhat reduced by a partial cancellation with additional contributions coming from the suppressed $A$-terms.) Given that the logarithmic enhancement is not large, $B \mu$ is typically smaller than $m_{H_d}^2$, favoring moderate-to-large values of $\tan \beta$.  That said, $m_{H_d}^2$ is significantly decreased by RG evolution, so that a wide range of $\tan \beta$ may be realized depending on the details of the soft spectrum.

Thus all the ingredients necessary for EWSB arise in the MSSM-like limit of the theory.  Radiative EWSB is possible thanks to the added RG contribution of the $h$ Yukawa, provided $g_M$ is not too large. The $\mu$ term arises supersymmetrically, and $B \mu$ is generated radiatively. 

More appealingly, the little hierarchy problem is alleviated by corrections to the tree-level Higgs mass.  As alluded to earlier, this limit naturally provides {\it two} corrections that lift the tree-level prediction for the Higgs mass and are significant for complementary values of $\tan \beta$. The first is the non-decoupling $D$-term mentioned in \Sec{sec:quartic}. The influence of this correction is maximal in the MSSM limit, where radiative corrections split $m_{H_u}^2$ from $m_{H_d}^2$ and lead to moderate or large values of $\tan \beta$. The second is the quartic superpotential correction in \Eq{eqn:mssmlim}. Such a quartic correction is a well-known means of raising the tree-level prediction for the lightest neutral Higgs mass. In this case, it shifts the Higgs mass by an amount \cite{Dine:2007xi}
\be
\delta m_h^2 \simeq \frac{8 m_{A^0}^2}{m_{A^0}^2 - m_Z^2}  \frac{\mu_N^2}{m_N^2} \frac{v^2}{\tan \beta},
\ee
where $m_{A^0}$ is the pseudoscalar mass.  This shift is particularly significant for $m_N \gtrsim \mu_N$ and moderate $\tan \beta$. Hence even in the MSSM-like limit of this model, there are significant additional contributions to the tree-level prediction for the Higgs mass, but they are qualitatively different than in the supersymmetric limit.

\section{Phenomenology}
 \label{sec:pheno}

Here, we will briefly discuss the main phenomenological features of our scenario, beyond the Higgs boson spectrum described in the previous section.  The collider phenomenology is similar to the MSSM or NMSSM, with the notable addition of TeV-scale exotic states.  

\subsection{SSM States}
\label{subsec:ssmstates}

As SUSY breaking occurs at around 100 TeV in our scenario, the gravitino is light, on the order of a few eV and safe from cosmological constraints \cite{Pagels:1981ke,Viel:2005qj}.  Thus, the leading phenomenology is that of a typical SSM with a gravitino as the lightest supersymmetric particle (LSP)  \cite{Dimopoulos:1996vz,Ambrosanio:1996zr,Dimopoulos:1996va,Ambrosanio:1996jn}.  The identity of the next-to-lightest supersymmetric particle (NLSP) depends on the details of the Higgs sector and the electroweak gaugino masses.  

While the singlino is often light  in many NMSSM models, that is not the case in the supersymmetric limit of EWSB discussed in \Sec{sec:EWSBsusy}. The singlino gets a large Dirac mass with the higgsinos of $\mathcal{O}(hv)$, and $h$ can be quite large.  In this region of parameter space, the lightest neutralino is often mostly bino, but its mass is sensitive to the relative sign of the gaugino and higgsino mass parameters.  In the basis where $h\langle N \rangle >0$,  if $m_{\lambda_1}>0$ then the lightest neutralino will be a few tens of GeV, but because it is mostly bino, it is still safe from limits on direct searches at colliders.  On the other hand, if $m_{\lambda_1}<0$ then the lightest neutralino and lightest chargino have similar masses, and the lightest chargino can often be the NLSP.  If electroweak symmetry is broken in the MSSM limit as in \Sec{sec:EWSBmssm}, then the NLSP can either be a neutralino (typically with a large bino and/or higgsino fraction) or a sfermion (typically a right- or left-handed stau).

The identity of the NLSP determines much of the collider phenomenology. Because the gravitino is light, the decay of the NLSP is prompt. For a mostly bino NLSP, SUSY events will often contain two photons, a classic signature of low-scale gauge mediation~\cite{Dimopoulos:1996vz,Ambrosanio:1996zr}. If kinematically allowed, there will be some events where the NLSP decays to a $Z$.  If the NLSP is a chargino, it is mostly higgsino but it still has some wino fraction, so the two-body decay to a $W$ and a gravitino, while suppressed by a mixing angle, will generally dominate over the three-body decay through a virtual Higgs to $b$ quarks.  A sfermion NLSP typically decays to its corresponding fermion via gravitino emission.

As discussed in \Sec{sec:MSSMsoft}, the soft masses for the MSSM states arise from gauge mediation, but with a number deviations from the minimal gauge mediation predictions.  First, the gaugino masses have a different parametric behavior than in minimal gauge mediation, because the gaugino masses arise from $R$-symmetry breaking.  Thus, the gauginos may be parametrically heavier or lighter than the sfermions depending on the details of the SUSY-breaking dynamics.  Second, the presence of additional heavy $SU(2)$ and $U(1)$ gauge bosons means that the soft spectrum is modified as in Higgsed gauge mediation \cite{Gorbatov:2008qa}.  Typically, the heavy $SU(2)$ gauge bosons only affect the Higgs spectrum since $g_F < g_M$ and the matter sfermions are only charged under $SU(2)_F$.  

The contribution from hypercharge is particularly interesting.  The hierarchy of $g_H$ and $g_V$ is a priori unknown, and since the matter sfermions  are charged under $U(1)_H$, the sfermions with larger hypercharge can be heavier than naively expected if $g_H > g_V$. Even when $g_H$ is only slightly larger than $g_V$, the large power of the ratio of the gauge couplings in \Eq{eqn:MatterSoft} raises the mass of the right-handed sleptons above the mass of the left-handed sleptons.  For larger values of the $U(1)_H$ coupling, $g_H \sim 1.5$, the mass of the right-handed sleptons is comparable to the squarks, which effectively decouples the right-handed sleptons at the LHC.\footnote{Needless to say, if either of the IR $U(1)$ gauge couplings are taken to be too large, there will be $U(1)$ Landau poles in the UV in addition to the potential $SU(3)_C$ Landau pole.}   In this regime the up-type right-handed squarks also become measurably heavier than the other squarks.   In certain cases, depending on the chargino and neutralino spectrum, the NLSP could even be a sneutrino.  

As long as the gluino and squarks are lighter than a few TeV, then the LHC will be able to probe the supersymmetric spectrum through cascade decays initiated by pair- or associated-production of the colored states.  Indeed, recent LHC searches for general gauge mediated models \cite{Meade:2008wd,Buican:2008ws} already constrain the parameter space with a bino-like NLSP \cite{Chatrchyan:2011wc}.  

\subsection{Colored Exotics}
\label{subsec:coloredexotics}

Beyond the SSM field content, there is novel phenomenology from the TeV-scale exotic states that participate in SUSY breaking.  Much of the additional matter charged under the SM obtains masses at---or one loop factor below---the scale of SUSY breaking, and are therefore too heavy to be produced at colliders like the LHC.  However, the fermionic components of the pseudo-modulus $X'$ obtain a mass of order $m_{\psi_{X'}} = (N_f -3) \gamma h^2  m_\Phi  \sim  \mathcal{O}({\rm TeV} )$ from $R$-symmetry breaking. These masses are parametrically of the same order as (though typically somewhat larger than) the SM gaugino masses, $m_\lambda \sim g_{SM}^2  m_\Phi (\mu_2/\mu_1)^4$.  As long as there is not a huge hierarchy between $\mu_1$ and $\mu_2$, the pseudo-moduli can play an interesting role in TeV-scale physics.\footnote{In addition, the pseudo-modulini masses are proportional to the parameter $\gamma$, which can be made small without changing the overall phenomenology.}

We will focus on colored exotic particles since these are the easiest to produce at the LHC.  The precise number and SM charges of these pseudo-modulini depend on how $SU(3)_C$ is embedded in the $SU(N_f - 3)$ flavor symmetry.  For the simple case of $N_f = 7, N_c = 5$, the pseudo-modulini amount to three fields $\psi_{X_3}$, $\psi_{\overline{X}_3}$, $\psi_{X'_8}$,  which respectively transform as $\mathbf{3}$, $\mathbf{\overline{3}}$, $\mathbf{8}$ under $SU(3)_C$. In the absence of any additional interactions (and ignoring the non-perturbative superpotential), these pseudo-modulini are all stable.  Thus, if they are produced, they will bind with SM quarks to form $R$-hadrons, some of which are electrically charged.  The lightest $R$-hadron associated with each pseudo-modulino is stable at the level of the IR superpotential.  (For a review of $R$-hadron phenomenology, see \Ref{Fairbairn:2006gg}.)

A cosmological population of stable $R$-hadrons is ruled out by bounds on heavy stable particles, thus the pseudo-modulini must be induced to decay.  To be conservative, we will arrange these decays to occur before big bang nucleosynthesis (BBN).  Pseudo-modulini decays may occur through dimension-five or dimension-six operators, arising due to symmetry-breaking interactions at a higher scale $M_*$.  There are a variety of scales in the theory at which such new physics may enter.  For example, there is the scale $\lambda h \Lambda$ of elementary fields coupled to $Q \overline Q$ (such as the spectators $S$); the scale $\Lambda_0$ at which the quartic operators $(Q \overline Q)^2$ are generated; or of course $M_{\rm GUT}$ or $M_{\rm Pl}$.  Given a new dimension-five operator generated at the scale $M_*$ (which may conceivably be any of the above scales, ranging from $10^5$--$10^{18}$ GeV), the lifetime of a pseudo-modulino is of order 
\be
\tau \simeq 8 \pi \frac{M_*^2}{m_{\psi_{X'}}^3} \simeq 2 \times 10^{-22} \, {\rm s} \,\left( \frac{\rm TeV}{m_{\psi_{X'}}}\right)^3 \left( \frac{M_*}{10^5 {\rm \, GeV}} \right)^2.
\ee
This leads comfortably to decays before BBN for all $M_* \lesssim M_{\rm GUT}$, and for $M_* \simeq M_{\rm GUT}$, late pseudo-modulino decays around $10^2$--$10^4$ seconds may even help to explain the cosmic lithium anomaly \cite{Jedamzik:2004er, Bailly:2008yy}.  For dimension-six decays,
\be
\tau \simeq 8 \pi \frac{M_*^4}{m_{\psi_{X'}}^5} \simeq 2 \times 10^{-18} \, {\rm s} \,\left( \frac{\rm TeV}{m_{\psi_{X'}}}\right)^5 \left( \frac{M_*}{10^5 {\rm \, GeV}} \right)^4,
\ee
for which $M_* \lesssim 10^{10}$ GeV is safe from BBN constraints.

Schematically, dimension-five UV operators which can lead to decays of $\psi_{X_3}$ and $\psi_{\overline{X}_3}$ are 
\be\label{eq:dim5}
W  = \frac{1}{M_*} Q \overline{Q} \Psi \Psi',
\ee
where $\Psi$ and $\Psi'$ are SM matter multiplets of the appropriate charge.  Below the confinement scale, these operators lead to decays of a pseudo-modulino to a fermion and a sfermion, such as
\be
\psi_{X_3}  \rightarrow q \tilde{\ell}, \qquad \psi_{\overline{X}_3} \rightarrow u^c \tilde{e}^c.
\ee
The octet $\psi_{X'_8}$ can only decay at dimension-six, through for example
\be
W = \frac{1}{M_*^2} \tr\left(Q \overline{Q} W^\alpha W_\alpha \right),
\ee
where $W_\alpha$ is for the SM color gauge field and the trace is over color indices.  Below the confinement scale, this leads to a transition color dipole decay to a gluon and a gluino
\be
\psi_{X'_8} \rightarrow g \tilde{g}.
\ee

These colored pseudo-modulini can be pair-produced via QCD processes.  For sufficiently large $M_*$, the pseudo-modulini can be stable on collider scales, leading to many of the spectacular collider signatures of $R$-hadrons (see \Ref{Khachatryan:2011ts,Aad:2011yf} for recent searches).  For example, a pseudo-modulino produced at colliders may travel at least 1 cm in the detector before decay provided $M_* \gtrsim 5 \times 10^{10}$ GeV (for dimension-five operators) or $M_* \gtrsim 7 \times 10^6$ GeV (for dimension-six operators).  For longer lifetimes, (charged) $R$-hadrons can stop in the calorimeters and undergo late decays in beam-off periods \cite{Arvanitaki:2005nq,Khachatryan:2010uf}. 

Finally, we note that some components  of the messengers arising from the $\{ \rho, Z \}$ sector enjoy a $\mathbb{Z}_2$ messenger parity and are thus stable at the level of the IR effective theory. Such stable messengers typically give rise to an overabundance of dark matter, which may be avoided by inducing decays via higher-dimension operators similar to \Eq{eq:dim5}.

\section{Conclusions}
\label{sec:conclude}

Using the power of duality, we have constructed a realistic composite Higgs theory where SUSY breaking and EWSB are intimately connected.  The dual magnetic gauge group plays a key role in our construction, with the Higgs multiplets arising as ``fat'' dual magnetic squarks.  This model naturally incorporates messengers for gauge mediation, which generates plausible SSM soft masses.  The physical Higgs boson is typically heavy owing to a combination of non-decoupling $D$-terms and a large NMSSM-like quartic coupling.

This theory showcases how the rich dynamics of SQCD can have a direct impact on electroweak scale physics.  The phenomenon of color-flavor locking is a key feature of metastable SUSY breaking in SQCD, but it plays an even more important role in this construction, since it allows for a large top Yukawa coupling even though the top quark is  elementary.  Quasi-stable pseudo-modulini are a generic feature of SQCD models, and they appear here as well, leading to long-lived colored states that may be kinematically accessible at the LHC.  While we have focused on the regime $\frac{3}{2}N_c > N_f > N_c$ where the theory is manifestly calculable, the special case of $N_f = 6$ and $N_c = 4$ can exhibit the desired phenomenology with a minimal set of particles, as discussed further in \App{app:64}.

After a year of successful data taking at the LHC, we are poised to understand the origin of EWSB.  While SUSY theories with elementary Higgs bosons have long been an attractive approach to the hierarchy problem, composite Higgs theories offer a plausible alternative.   We find SUSY composite Higgs bosons to be particularly appealing given the difficulty of achieving a sufficiently heavy physical Higgs boson in the MSSM alone, and we are encouraged by the relative simplicity and calculability of our proposed scenario.  Ultimately, the LHC will provide insight into whether there truly is a desert above the electroweak scale, or whether EWSB is only a small part of rich short-distance dynamics. \\

\noindent {\bf Note added:} While this paper was in preparation, we learned of \Ref{JohnCsaba}, which also envisions a novel use of the dual magnetic gauge group.

\section*{Acknowledgements}

We would like to thank Csaba Csaki, Daniel Green, Andrey Katz, Markus Luty, Yasunori Nomura, Matt Strassler, and John Terning for helpful discussions. NC and DS would like to acknowledge the hospitality of the MIT Center for Theoretical Physics, where this work was initiated.  DS would like to thank the Berkeley Center for Theoretical Physics for their hospitality.  JT would like to thank the UMN Fine Theoretical Physics Institute for their hospitality while this work was being completed.  NC is supported in part by the NSF under grant PHY-0907744 and gratefully acknowledges support from the Institute for Advanced Study.  DS is supported in part by the NSF under grant PHY-0910467 and gratefully acknowledges support from the Maryland Center for Fundamental Physics.   JT is supported by the U.S. Department of Energy under cooperative research agreement Contract Number DE-FG02-05ER41360.

\appendix

\section{The Theory with $N_f = 7, N_c = 5$ \label{app:75}}

\TABLE[t]{
\begin{tabular}{|c|c|cccc|} \hline
 & $SU(5)$ & $SU(3)_C$ & $SU(2)_F$ & $U(1)_V$ & $U(1)_H$  \\ \hline
$P$ & ${\bf 5}$ & ${\bf 1}$ & ${\bf 1}$ & $-1/3$ & $-1/6$ \\
$\overline{P}$ & ${\bf \overline 5}$ & ${\bf 1}$ & ${\bf 1}$ & $+1/3$ & $+1/6$ \\
$Q_2$& ${\bf 5}$ & ${\bf 1}$ & ${\bf 2}$ & $-1/3$ & $+1/3$ \\
$\overline{Q}_2$& ${\bf \overline 5}$ & ${\bf 1}$ & ${\bf  2}$ &  $+1/3$ &  $-1/3$ \\
$Q_3$& ${\bf 5}$ & ${\bf \overline 3 }$ & ${\bf 1}$ & $-1/3$ & $+1/6$ \\
$\overline{Q}_3$& ${\bf \overline 5}$ & ${\bf 3 }$ & ${\bf 1}$ & $+1/3$ & $-1/6$ \\
$Q_1$& ${\bf 5}$ & ${\bf 1}$ & ${\bf 1}$ & $-1/3$ & $-1/6$ \\
$\overline{Q}_1$& ${\bf \overline 5}$ & ${\bf  1}$ & ${\bf 1}$ & $+1/3$ & $+1/6$ \\
 \hline
$S_2$ & ${\bf 1}$ & ${\bf 1}$ & ${\bf 2}$ & 0 & $-1/2$ \\
$\overline{S}_2$ & ${\bf 1}$ & ${\bf 1}$ & ${\bf 2}$ & 0  & $+1/2$ \\
$S_3$ & ${\bf 1}$ & ${\bf 3}$ & ${\bf 1}$ & 0 & $-1/3$ \\
$\overline{S}_3$ & ${\bf 1}$ & ${\bf \overline 3}$ & ${\bf 1}$ & 0 & $+1/3$ \\
$S_1$ & ${\bf 1}$ & ${\bf 1}$ & ${\bf 1}$ & 0 & 0 \\
$\overline{S}_1$ & ${\bf 1}$ & ${\bf 1}$ & ${\bf 1}$ & 0 & 0 \\
\hline
$\CO^{u}_{ij}$ & ${\bf 1}$ & ${\bf 1}$ & ${\bf 2}$ & 0 & $-1/2$ \\
$\CO^d_{ij}$ & ${\bf 1}$ & ${\bf 1}$ & ${\bf 2}$ & 0 & $+1/2$  \\ \hline
\end{tabular}
\caption{The full UV field content for $N_f = 7, N_c =5$.  Note that $U(1)_V$ is proportional to electric antiquark number. \label{tab:uvfields75}}
}

\TABLE[t]{
\begin{tabular}{|c|c|cccc|} \hline
 & $SU(2)_M$ & $SU(3)_C$ & $SU(2)_F$ & $U(1)_V$ & $U(1)_H$  \\ \hline
$H_u$ & ${\bf 2}$ & ${\bf 1}$ & ${\bf 1}$ & $+1/3$ & $+1/6$ \\
$H_d$ & ${\bf 2}$ & ${\bf 1}$ & ${\bf 1}$ & $-1/3$ & $-1/6$ \\
$\sigma$  & ${\bf 2}$ & ${\bf 1}$ & ${\bf 2}$ & $+1/3$ & $-1/3$  \\
$\overline{\sigma}$  & ${\bf 2}$ & ${\bf 1}$ & ${\bf 2}$ & $-1/3$ & $+1/3$  \\
$\rho_3$ & ${\bf 2}$ & ${\bf 3}$ & ${\bf 1}$ & $+1/3$ & $-1/6$ \\
$\overline{\rho}_3$ & ${\bf 2}$ & ${\bf \overline 3}$ & ${\bf 1}$ & $-1/3$ & $+1/6$ \\
$\rho_1$ & ${\bf 2}$ & ${\bf 1}$ & ${\bf 1}$ & $+1/3$ & $+1/6$ \\ 
$\overline{\rho}_1$ & ${\bf 2}$ & ${\bf 1}$ & ${\bf 1}$ & $-1/3$ & $-1/6$ \\ 
 \hline
$N$ & ${\bf 1}$ & ${\bf 1}$ & ${\bf 1}$ & 0 & 0 \\
$Y$ & ${\bf 1}$ & ${\bf 1}$ & ${\bf 3} \oplus {\bf 1}$ & 0 & 0 \\
$Z_3$ & ${\bf 1}$ & ${\bf \overline 3}$ & ${\bf 2}$ & 0 & $-1/6$ \\
$\overline{Z}_3$ & ${\bf 1}$ & ${\bf 3}$ & ${\bf 2}$ & 0 & $+1/6$ \\
$Z_1$ & ${\bf 1}$ & ${\bf 1}$ & ${\bf 2}$ & 0 & $-1/2$ \\
$\overline{Z}_1$ & ${\bf 1}$ & ${\bf 1}$ & ${\bf 2}$ & 0 & $+1/2$ \\
$X_8$ & ${\bf 1}$ & ${\bf 8} \oplus {\bf 1}$ & ${\bf 1} $ & 0 & 0 \\
$X_1$ & ${\bf 1}$ & ${\bf 1}$ & ${\bf 1} $ & 0 & 0 \\  
$X_3$ & ${\bf 1}$ & ${\bf 3}$ & ${\bf 1} $ & 0 & $-1/3$ \\
$\overline{X}_3$ & ${\bf 1}$ & ${\bf \overline 3}$ & ${\bf 1} $ & 0 & $+1/3$  \\
$\Sigma_2$ & ${\bf 1}$ & ${\bf 1}$ & ${\bf 2}$ & 0 & $+1/2$  \\
$\overline{\Sigma}_2$& ${\bf 1}$ & ${\bf 1}$ & ${\bf 2}$ & 0 & $-1/2$  \\
$\Sigma_3$ & ${\bf 1}$ & ${\bf \overline 3}$ & ${\bf 1}$ & 0 & $+1/3$  \\
$\overline{\Sigma}_3$ & ${\bf 1}$ & ${\bf 3}$ & ${\bf 1}$ & 0 & $-1/3$  \\
$\Sigma_1$ & ${\bf 1}$ & ${\bf 1}$ & ${\bf 1}$ & 0 & 0  \\
$\overline{\Sigma}_1$ & ${\bf 1}$ & ${\bf 1}$ & ${\bf 1}$ & 0 & 0  \\ \hline
\end{tabular}
\caption{The IR field content for $N_f = 7, N_c =5$.  Note that $U(1)_V$ is proportional to magnetic quark number.  The spectator fields $S$, $\overline{S}$ and the SM chiral operators $\mathcal{O}^u$,  $\mathcal{O}^d$ are given in \Tab{tab:uvfields75}. \label{tab:irfields75}}
}

In this appendix, we provide a detailed accounting of the states and charge assignments for the minimal theory with $N_f = 7, N_c = 5$. This is useful for determining the appropriate $U(1)$ charge assignments, and hence the corresponding SM charges of various TeV-scale exotics.  We use the notation of previous sections but decompose the fields according to their $SU(2)_F \times SU(3)_C$ charges.  For a fundamental $Q$ of $SU(N_f - 1)_D$, we use the subscript notation
\be
Q = \left(\begin{array}{c} Q_2 \\ Q_3 \\ Q_1  \end{array}  \right).
\ee
For an adjoint plus singlet of $SU(N_f - 3)_D$, we use
\be
X = \left(\begin{array}{cc} X_8 & \overline{X}_3 \\ X_ 3 & X_1  \end{array}  \right).
\ee
We collect the fields of the UV theory in \Tab{tab:uvfields75} and those of the IR theory in \Tab{tab:irfields75}.

\section{A Novel Theory with $N_f = 6, N_c = 4$ \label{app:64}}

Thus far, we have taken care to avoid the case of $N_f = 6, N_c = 4$, for which $N_f = \frac{3}{2} N_c$. This theory still possesses an IR free dual description, as the IR gauge coupling runs free at two loops. However, it also possesses a {\it marginal} nonperturbative superpotential of the form
\be
W_{\rm NP} = 2 (h^6 N \det \Phi)^{1/2}.
\ee
Given that the superpotential is marginal, it is natural to worry that there are no metastable nonsupersymmetric vacua, since additional $F$-terms may be cancelled off at tree level by giving vevs to $N$ and $\Phi$. While it is certainly the case that this nonperturbative superpotential leads to additional supersymmetric vacua, we will argue that these vacua may be made parametrically distant from the metastable nonsupersymmetric vacuum. The key feature is that the hierarchy of scales between the SUSY-breaking and EWSB sectors pushes these supersymmetric vacua out to large field values and allows for local nonsupersymmetric vacua, much as in the case for dynamical SUSY breaking with a quadratic superpotential deformation \cite{Essig:2008kz}.

This vacuum structure is attractive for two reasons. The first is that it leads to spontaneous $R$-symmetry breaking in the SUSY-breaking sector without introducing any further quadratic deformations of the SUSY-breaking fields; this is simply because the $R$-breaking nonperturbative superpotential in this theory is marginal and hence induces $R$-symmetry breaking in the local vacuum.\footnote{In the analysis below, we will retain one $R$-breaking term ($m_N$) as a calculational handle.}  The second is that this renders the theory with $N_f = 6, N_c = 4$ viable for SUSY breaking and EWSB (in a limited sense, which we will make clear shortly). This theory possesses sufficiently few additional fields charged under the SM that there are no Landau poles in the SM gauge couplings up to the Planck scale. However, the tradeoff of exploiting this vacuum structure is that EWSB is entirely MSSM-like (albeit still with a large correction to the Higgs quartic). 

For simplicity, let us first consider the case where $m_N$ is large ($\gg \mu_1, \mu_2$) but the hierarchy between $m_N$ and $\mu_N$ is such that $\mu_H \equiv \mu_N^2 / m_N \sim m_W$.  This corresponds to a mass term for the magnetic quarks $H_u, H_d$.  Integrating out $N$ at the scale $m_N$, the IR theory has a dynamical superpotential
\be
W_{\rm NP} = 2 (h^5 \mu_H \det \Phi)^{1/2}.
\ee
Adding this relevant contribution to our IR superpotential leads to a supersymmetric vacuum at $\Phi \sim \frac{1}{h} \mu_2 (\mu_2/\mu_H)^{1/3}$. However, we will now show that there is also a metastable nonsupersymmetric vacuum closer to the origin. 

We can determine the location of this vacuum by turning on a nonsupersymmetric CW mass for $\Phi$ in the scalar potential, $\sim m_{\CW}^2 |\Phi|^2,$ originating from the nonzero $F$-terms in our SUSY-breaking sector. A metastable vacuum arises from the competition between this soft mass and the leading tachyonic term for $\Phi$. The ensuing minimum of the potential lies at
\be
\Phi \sim \frac{1}{h} \mu_H \left( \frac{h^2 \mu_2}{m_{\CW}} \right)^4.
\ee
Since $m_{\CW}$ is suppressed by a loop factor $b \equiv \frac{\log 4 - 1}{4 \pi^2}$ relative to $\mu_2$ (recall $m_{\CW} = \sqrt{b} h^2 \mu_2$), this means the potential turns around only at large field values. For this analysis to be correct, we require first
\be
\mu_H \left(\frac{h^2 \mu_2}{ m_{\CW}} \right)^4 \lesssim h^2 \mu_2.
\ee
This is the condition that the CW contribution is still significant where the potential turns around, since the CW potential falls off once we go above the mass scale of the fields running in the loop. This amounts to requiring $\mu_H \leq h^2 b^2 \mu_2.$ We also require the minimum to lie closer to the origin than the SUSY minimum. This condition corresponds to 
\be
\mu_H < \frac{1}{h^2} \frac{m_{\CW}^3}{h \mu_2^2},
\ee
and hence $\mu_H < b^{3/2} \mu_2.$ This is typically a weaker condition than the CW condition for $h \sim 1$.

We must also verify that the magnetic quarks are not tachyonic in this vacuum; indeed, their masses are positive in the vicinity of $\mu_H \sim h^2 b^2 \mu_2,$ so the vacuum is metastable as desired. The vev of $\Phi$ in this vacuum is $\langle \Phi \rangle \sim h \mu_2$ which gives a large spontaneous $R$-symmetry breaking. This results in gaugino masses that are the same size as the scalar soft masses.

Finally, we must verify that the metastable vacuum is sufficiently long-lived. The bounce action for tunneling into the supersymmetric vacuum in the square approximation is
\be
B \sim 2\pi^2 \frac{1}{h^6} \left(\frac{\mu_2}{\mu_H} \right)^{4/3} \sim \frac{2 \pi^2}{b^{8/3}},
\ee
which is parametrically large and suppresses tunneling exponentially.  Thus, the lifetime of the metastable vacuum is significantly longer than the age of the universe.

This analysis required $m_N$ sufficiently large that we could integrate out $N$ and neglect its dynamics in the local vacuum. While this limit is perfectly valid, it removes the NMSSM-like features of the theory at a high energy, leading merely to an MSSM-like Higgs sector with modified $D$-terms in the IR. One might hope that this attractive vacuum structure might persist for smaller values of $m_N$. However, while there is still a metastable vacuum parametrically distant from the supersymmetric vacua in this limit, the vacuum energy is minimized by giving an $\CO(h \mu_2)$ vev to $N$. This vev leads to a $\mu$-term that is far too large for natural EWSB. Hence viable SUSY breaking and EWSB in the theory with $N_f = 6$ and $N_c = 4$ requires large values of $m_N$, leaving only the non-decoupling $D$-term as a low-energy signature of the UV dynamics.

\section{Spectrum for General $N_f$ \label{app:soft}}

Given the intertwined dynamics of duality, SUSY breaking, and EWSB, it is useful to account for the full spectrum of dynamical fields and their masses for general $N_f$. Here, we have organized the fields by their transformation properties under $SU(2)_L \times SU(N_f - 3)_D$; SM charges may be obtained by decomposing representations accordingly after gauging $SU(3)_C \subset SU(N_f - 3)_D$ and as well as $U(1)_V$ and $U(1)_H$.  The fermionic spectrum is listed in \Tab{tab:fermions}, while the bosonic spectrum is listed in \Tab{tab:bosons}.  In addition to the matter listed in these tables arising from the chiral multiplets of the theory, there are heavy vector multiplets of $(SU(2)_M \times SU(2)_F)/SU(2)_L$ with mass $\sqrt{2 g_M^2 + 2 g_F^2} \mu_1$ and of $(U(1)_H \times U(1)_V)/U(1)_Y$ with mass $\sqrt{2 g_V^2 + 2 g_H^2} \mu_1.$ The orthogonal vector multiplets of $SU(2)_L \times U(1)_Y$ are massless prior to electroweak symmetry breaking.

\TABLE[t]{
\begin{tabular}{|c|cccc|} \hline
& Weyl d.o.f. & mass & $SU(N_f - 3)_D$ & $SU(2)_L$ \\ \hline
$\tr X$ & 1 & 0 & {\bf 1} & {\bf 1}  \\  \hline
$X'$ & $(N_f - 3)^2-1$ & $(N_f - 3) \gamma h^2  m_\Phi$ & {\bf Adj.} & {\bf 1} \\ \hline
 & 4 & $\CO(h \mu_1)$ & {\bf 1} & {\bf 3+1} \\ 
$Y,$ & 4 & $\CO(h \mu_1)$ & {\bf 1} & {\bf 3+1}  \\ 
$\sigma, \overline{\sigma}$& 3 & $\sqrt{2g_M^2 + 2g_F^2} \mu_1$ & {\bf 1} & {\bf 3 }  \\
& 1 & $ \sqrt{2 g_V^2 + 2 g_H^2} \mu_1$ & {\bf 1} & {\bf 1} \\ \hline
$Z, \overline{Z},$ & $4 (N_f - 3)$ & $\CO(h \mu_1)$ & $\Box + \overline \Box$ & ${\bf 2 + 2}$  \\
$\rho, \overline{\rho}$ & $4 (N_f - 3)$ & $\CO(h \mu_1)$ &  $\Box + \overline \Box$ & ${\bf 2 + 2}$  \\ \hline
$N$ & 1 & $\CO(m_W)$ &  {\bf 1} &  {\bf 1}   \\ \hline
$H_u, H_d$ & 4 & $\CO(m_W)$ &  {\bf 1}  &  ${\bf 2 + 2}$  \\ \hline 
$\Sigma, \overline \Sigma,$ & $2 (N_f -1)$ & $\lambda h \Lambda$ & $\Box + \overline \Box$& ${\bf 2 +2}$ \\
$S, \overline S$ & $2 (N_f -1)$ & $\lambda h \Lambda$ & $\Box + \overline \Box$& ${\bf 2 +2}$ \\  \hline
\end{tabular}
\caption{Fermionic field content for general $N_f$. The mass eigenstates on the right side of the table are linear combinations of the gauge eigenstates listed on the left side.  Some of the $\sigma$, $\overline{\sigma}$ modes get Dirac masses with the heavy $SU(2)$ and $U(1)$ gauginos. \label{tab:fermions}}}

\TABLE[t]{
\begin{tabular}{|c|cccc|} \hline
& Real d.o.f. & mass & $SU(N_f - 3)_D$ & $SU(2)_L$ \\ \hline
$\tr X$ & 1 & $\CO(m_{\CW})$ & {\bf 1} & {\bf 1}   \\  
& 1 & $\CO(\sqrt{h^3 m_\Phi \mu_2^2/\langle X \rangle})$ & {\bf 1} & {\bf 1} \\ \hline
$X'$ & $2 (N_f - 3)^2-2$ & $\CO(m_{\CW})$ & {\bf Adj.} & {\bf 1} \\ \hline
 & 8 & $\CO(h \mu_1)$ & {\bf 1} & {\bf 3+1} \\ 
$Y,$ & 8 & $\CO(h \mu_1)$ & {\bf 1} & {\bf 3+1}  \\ 
$\sigma, \overline{\sigma}$& 6 & $\CO(\sqrt{2 g_M^2 + 2 g_F^2} \mu_1)$ & {\bf 1} & {\bf 3}  \\
& 2 & $ \CO(\sqrt{2 g_V^2+2 g_H^2} \mu_1)$ & {\bf 1} & {\bf 1} \\ \hline
$Z, \overline Z,$ & $8 (N_f - 3)$ & $\CO(h \mu_1)$ & $\Box + \overline \Box$ & ${\bf 2 + 2}$  \\
$\rho, \overline \rho$ & $4 (N_f - 3)$ & $\CO(h \mu_1)$ &  $\Box + \overline \Box$ & ${\bf 2 + 2}$  \\ 
&$4 (N_f - 3)$ & $\CO(h \mu_1)$ & $\Box + \overline \Box$ & ${\bf 2 + 2}$ \\ \hline
$N$ & 2 & $\CO(m_W)$ &  {\bf 1} &  {\bf 1}   \\ \hline
$H_u, H_d$ & 8 & $\CO(m_W)$ &  {\bf 1}  &  ${\bf 2 + 2}$  \\ \hline 
$\Sigma, \overline \Sigma,$ & $4 (N_f -1)$ & $\lambda h \Lambda$ & $\Box + \overline \Box$& ${\bf 2 +2}$ \\
$S, \overline S$ & $4 (N_f -1)$ & $\lambda h \Lambda$ & $\Box + \overline \Box$& ${\bf 2 +2}$ \\  \hline
\end{tabular}
\caption{Bosonic field content for general $N_f$. The mass eigenstates on the right side of the table are linear combinations of the gauge eigenstates listed on the left side. The representations in the last two columns correspond to complex scalars, with the exception of the pseudo-goldstones and sgoldstones of $\rho, \bar \rho$, which correspond to real scalars. Some of the $\sigma$, $\overline{\sigma}$ modes get masses because they are eaten by the heavy $SU(2)$ and $U(1)$ gauge bosons.  \label{tab:bosons}}}

\bibliography{FHbib}
\bibliographystyle{JHEP}

\end{document}